\documentclass[aps,prc,reprint,groupedaddress,onecolumn]{revtex4}
\usepackage{amsmath}
\usepackage{amssymb}
\usepackage{amsfonts}
\usepackage{graphics}
\usepackage[dvips]{color}
\def\beq{\begin{equation}}
\def\eeq{\end{equation}}
\def\beqarray{\begin{eqnarray}}
\def\eeqarray{\end{eqnarray}}

\begin{document}

\title{Wavelet representation of light-front quantum field theory}

\author{W.~N.~Polyzou}
\affiliation{Department of Physics and Astronomy, The University of
Iowa, Iowa City, IA 52242}

\date{\today}


\begin{abstract}  

  A formally exact discrete multi-resolution representation of quantum
  field theory on a light front is presented.  The formulation uses an
  orthonormal basis of compactly supported wavelets to expand the
  fields restricted to a light front.  The representation has a number
  of useful properties.  First, light-front preserving Poincar\'e
  transformations can be computed by transforming the arguments of the
  basis functions.  The discrete field operators, which are defined by
  integrating the product of the field and a basis function over
  the light front, represent localized degrees of freedom on the
  light-front hyperplane.  These discrete fields are irreducible and
  the vacuum is formally trivial.  The light-front Hamiltonian and all
  of the Poincar\'e generators are linear combinations of normal
  ordered products of the discrete field operators with analytically
  computable constant coefficients.  The representation is discrete
  and has natural resolution and volume truncations like lattice
  formulations.  Because it is formally exact it is possible to
  systematically compute corrections for eliminated degrees of
  freedom.

\end{abstract}

\maketitle


\section{Introduction}

A discrete multi-resolution representation of quantum field theory on
a light front is presented.  Light-front formulations of quantum field
theory have advantages for calculating electroweak current matrix
elements in strongly interacting states in different frames.  Lattice
truncations have proved to be the most reliable method for
non-perturbative calculations of strongly interacting states, but
Lorentz transformation and scattering calculations are not naturally
formulated in the lattice representation.  The purpose of this work is to
investigate a representation of quantum field theory that has some of
the advantages of both approaches, although this initial work is
limited to canonical field theory rather than gauge theories.

In 1939 Wigner \cite{Wigner:1939cj} showed that the independence of
quantum observables in different inertial reference frames related by
Lorentz transformations and space-time translations requires the
existence of a dynamical unitary representation of Poincar\'e group on
the Hilbert space of the quantum theory.  Because there are many
independent paths to the future, consistency of the initial value
problem requires that a minimum of three of the infinitesimal
generators of the Poincar\'e group are interaction dependent.  In 1949
P. A. M. Dirac \cite{Dirac:1949cp} introduced three ``forms of
relativistic dynamics'' that are characterized by having the largest
interaction-independent subgroups.

Dirac's `front-form dynamics' is the only form of dynamics with the
minimal number, 3, of dynamical Poincar\'e generators.  The
interaction-independent subgroup is the seven-parameter subgroup that
leaves the hyperplane,
\beq
x^+ = x^0+ \hat{\mathbf{n}}\cdot \mathbf{x} =0
\label{i.1}
\eeq
invariant.  The light-front representation of quantum dynamics has
several advantages.  One is that the kinematic
(interaction-independent) subgroup has a three-parameter subgroup of
Lorentz boosts.  The subgroup property means that there are no
Wigner rotations for light-front boosts.  A consequence is that the magnetic
quantum numbers of the light-front spin are invariant with respect to
these boosts.  A second advantage is that the boosts are independent
of interactions.  This means that boosts can be computed by applying
the inverse transform to non-interacting basis states.  These
properties simplify theoretical treatments of electroweak
probes of strongly-interacting systems, where the initial and final
hadronic states are in different Lorentz frames.

In light-front quantum field theory 
\cite{fubibi:1965}
\cite{weinberg:1966}
\cite{susskind:1968}
\cite{bardakci:1968}
\cite{chang:1969}
\cite{soperL1970}
\cite{Leutwyler:1970wn}
\cite{Schlieder:1972qr}
\cite{Chang:1972xt}
\cite{coester:1992}
\cite{wilson:1994}
\cite{Brodsky:1998} there are additional advantages. 
These are consequences of the spectrum of the generator
\beq
p^+ = p^0+ \hat{\mathbf{n}}\cdot \mathbf{p} \geq 0
\label{1.2}
\eeq
of translations in the
\beq
x^- = x^0- \hat{\mathbf{n}}\cdot \mathbf{x}
\label{1.3}
\eeq
direction tangent to the light-front.  The first property is that free
fields restricted to the light-front are irreducible.  This means that
both the creation and annihilation operators for a free field can be
constructed from the field restricted to the light front. It follows
that any operator on the free field Fock space can be expressed as a
function of free fields restricted to the light front.  The second
advantage is that interactions that commute with the
interaction-independent subgroup leave the Fock vacuum invariant.
This means that it is possible to express all of the Poincar\'e
generators as operators on the free-field Fock space.  There are
ultraviolet and infrared ($p^+=0$) singularities in the light-front
Hamiltonian due to local operator products, which could impact these
properties, however in an effective theory with ultraviolet and
infrared cutoffs the interaction still leaves the Fock vacuum
invariant and the light-front Hamiltonian can still be represented as
a function of the free fields on the light front.

Having an explicit vacuum along with an expression for the light-front
Hamiltonian,
\beq
P^- = P^0-\mathbf{P} \cdot \mathbf{n}, 
\label{1.3}
\eeq
in terms of the algebra of fields on the light front means
that it is possible to perform non-perturbative calculations by
diagonalizing the light-front Hamiltonian in the light-front
Fock space.

In a given experiment there is a relevant volume and a finite amount
of available energy.  The available energy limits the resolution of
the accessible degrees of freedom.  The number of degrees of freedom
with the limiting resolution that fit in the experimental volume is
finite.  It follows that it should be possible to accurately calculate
experimental observables using only these degrees of freedom.

Wavelets can be used to represent fields on the light front as linear
combinations of discrete field operators with different resolutions.
This provides a natural representation to make both volume and
resolution truncations consistent with a given reaction.  In addition
the representation is discrete, which is a natural representation for
computations.  Finally the basis functions are self-similar, so
truncations with different resolutions have a similar form.

There are many different types of wavelets that have been discussed in
the context of quantum field theory
\cite{best:1994}\cite{federbush:1995}\cite{halliday}\cite{Battle:1999}
\cite{best:2000}\cite{Ismail1:2003}\cite{Ismail2:2003}\cite{altaisky:2007}
\cite{albeverio:2009}\cite{altaisky:2010}
\cite{fatih2}\cite{altaisky:2013}\cite{altaisky:2013b}\cite{fatih1}
\cite{Brennen}\cite{altaisky:2016b}\cite{altaisky:2016c}\cite{Evenbly}
\cite{neuberger:1}\cite{tracie}\cite{altaisky:2017}.
The common feature is that the different functions have a common
structure related by translations
and scale transformations.  This work uses Daubechies' wavelets
\cite{daubechies:1988}\cite{daubechies}\cite{kaiser}\cite{resnikoff}\cite{jorgensen}.  
These
have the property that they are an orthonormal basis of functions with
compact support.  The price paid for the compact support is that they
have a limited smoothness.  It is also possible to use a wavelet basis
of Schwartz functions that are infinitely differentiable, but these
functions do not have compact support.

This work is an extension to the light front of the wavelet
representations of quantum field theory used in references
\cite{fatih1} \cite{fatih2} \cite{tracie}.  The notation and
development of the wavelet bases is identical to the development in
these references.  The difference is that the algebra generated by the
discrete fields and conjugate generalized momenta
in these papers is replaced by the irreducible algebra of fields on
a light front.  The light-front representation is formally exact and
has all of the advantages of any other representation of
light-front field theories.  

There are several motivations for considering this approach.  These
include
\begin{itemize}

\item[1.] Volume and resolution truncations can be performed naturally, the
resulting truncated theory is similar to a lattice truncation
\cite{PhysRevD.10.2445}\cite{Wilson:pr1976}, in the sense that it is a
theory involving a finite number of discrete degrees of freedom
associated with a given volume and resolution.

\item[2.] While the degrees of freedom are discrete, the field
  operators have a continuous space-time dependence.
  Kinematic Lorentz transformations can be computed by
  transforming the arguments of the basis functions.  While
  truncations necessarily break kinematic Lorentz invariance,
  kinematic Lorentz transformations can still be approximated by
  transforming the arguments of the basis functions.

\item[3.] Even though some truncations may lead to states
with energy below the Fock vacuum energy, the error in using the
free Fock vacuum as the lowest mass state of the truncated theory is due to
corrections that arise from the discarded degrees of freedom.

\item[4.] Since the representation is formally exact and $x^+$ is a
  continuous variable, there is a formulation of
  Haag-Ruelle \cite{Haag:1958vt}\cite{Ruelle:1962} \cite{jost}
  scattering in this representation.  Approximation methods need
  to be developed in the presence of truncations.
  
\end{itemize}

Some of the possible applications of the wavelet representation were
discussed in \cite{fatih2} in the context of canonical field theory.
There are a number of applications involving free fields that are
straightforward and should be instructive.  The advantage of free
fields is that they can be solved and used as a testing ground in
order to get an initial understanding of the convergence of truncated
theories.  One such application is understanding the restoration of
Poincar\'e invariance in truncated theories as the resolution is
improved.  An advantage of the wavelet representation is that this can
be checked locally, i.e. in a small volume \cite{fatih2}.
Understanding the restoration of Lorentz invariance is important for
approximating current matrix elements.  Another application involving
free fields is to test the convergence of free field commutator functions
or Wightman functions based on truncated fields to the exact
expressions.  These can be approximated by iterating the Heisenberg
field equations, which are simple in the free field case.  This could
provide some insight into the nature of convergence in interacting
theories.  In \cite{tracie} flow equation methods were used block
diagonalize the Hilbert space of a truncated free field theory by
resolution, constructing an effective Hamiltonian that involves only
coarse-scale degrees of freedom, but includes the dynamics of the
eliminated degrees of freedom.  This calculation provided some insight
into the complementary roles played by volume and resolution
truncations.

While the elementary calculations discussed above can provide insight
into the nature of approximations, the long-term goal is to use the
wavelet representation to perform calculations of observables in $3+1$
dimensional field theories.  Calculations in 3+1 dimensions are
considerably more complicated for interacting theories.  One
computational method is to use the fields to construct a basis by
applying discrete operators to the vacuum, and then diagonalizing the
light-front Hamiltonian in that basis.  The light-front representation
has the advantage that it is not necessary to first solve the vacuum
problem.  This method should be useful for modeling composite states
that are spatially localized.  This Hamiltonian approach is in the
same spirit as the basis light front quantization approach used in
\cite{james}.  Variational methods could also be employed for
low-lying composite states.  Another method that takes advantage of the
discrete nature of the wavelet representation is to use the
light-front Heisenberg equations to generate an expansion of the field
as a linear combination of products of fields restricted to the light
front.  Correlation functions can be computed by evaluating products
of these fields in the light-front vacuum. In this case while the
algebra is discrete, the number of terms grows with each iteration.  One
of the advantage of the wavelet representation is that interactions
involving different modes are self-similar and differ only by
multiplicative scaling coefficients.  A detailed study of the scaling
properties could help to formulate efficient approximations to the
solution of the light-front Heisenberg field equations by eliminating
irrelevant degrees of freedom.  Another potential use of the wavelet
representation would be in quantum computing.  In the wavelet
representation the field is replaced by discrete modes that only
interact locally.  This allows evolution over short time steps to be
represented by quantum circuits involving products of local
interactions.

This paper consists of thirteen sections.  The next section introduces 
the notation that will be used in this work, defines the light-front
kinematic subgroup and the Poincar\'e generators that generate
both the kinematic and dynamical Poincar\'e transformations.
Section three discusses the irreducibility of free fields on the
light front and properties of kinematically invariant interactions.
Section four discusses the structure of Poincar\'e generators on the
light front using Noether's theorem.  The wavelet basis is
constructed from the fixed point of a renormalization group equation in
section five.  Wavelet representations of fields restricted to the
light front are defined in section six.  Section seven has a short
discussion on kinematic Poincar\'e transformations of the fields in the
light-front representation.  In section eight the irreducibility of the
light-front free-field algebra and the triviality of the light-front
vacuum are used to construct vectors in the
light-front Fock space.  Dynamical equations in the light-front wavelet
representation are discussed in section nine. Dynamical computations
require expressions for the commutator of discrete fields on the
light front, which are computed in section ten.  In section eleven
the coefficients of the expansion of all ten Poincar\'e generators
as polynomials of discrete fields on the light front are computed.
Section twelve discusses truncations and section thirteen gives a summary
and outlook.

  
\section{Notation}

The light front is a three-dimensional hyperplane that is tangent
to the light cone.  It is defined by the constraint
\beq
x^+ := x^0 +\hat{\mathbf{n}} \cdot \mathbf{x} =0 .
\label{n.1}
\eeq
It is natural to introduce light-front coordinates of the 
four-vector $x^{\mu}$:
\beq
x^{\pm} := x^0 \pm\hat{\mathbf{n}} \cdot \mathbf{x}, \qquad
\mathbf{x}_{\perp} =  \hat{\mathbf{n}} \times ({\mathbf{x}} \times
\hat{\mathbf{n}}).
\label{n.2}
\eeq
The components
\beq
\tilde{\mathbf{x}} := (x^-,\mathbf{x}_{\perp})
\label{n.3}
\eeq
are coordinates of points on the light-front hyperplane.  These will be
referred to as light-front 3-vectors.  In what follows the light front
defined by $\hat{\mathbf{n}}=\hat{\mathbf{z}}$ will be used. 

The contravariant light-front components are
\beq
x_{\pm}  = -x^{\mp} \qquad x_{i \perp}= x_{\perp}^i 
\label{n.4}
\eeq
and the Lorentz-invariant scalar product of two light-front vectors is
\beq
x \cdot y:=
-{1 \over 2} x^+ y^-
-{1 \over 2} x^- y^+ + \mathbf{x}_{\perp}\cdot \mathbf{y}_{\perp}  =
{1 \over 2} (x^+ y_+ + x^- y_-) + x^1 y_1 + x^2 y_2 .
\label{n.5}
\eeq
For computational purposes it is useful to represent four vectors
by $2 \times 2$ Hermitian matrices.  The coordinate matrix is
constructed by contracting the four vector $x^{\mu}$ with the Pauli
matrices and the identity:
\beq
X = x^{\mu}\sigma_{\mu} = 
\left (
\begin{array}{cc}
x^+ & x_{\perp}^*\\
x_{\perp} & x^- \\
\end{array}
\right )
\qquad
x^{\mu} = {1 \over 2}\mbox{Tr}(\sigma_{\mu} X)
\quad
x_{\perp} = x^1 + i x^2. 
\label{n.6}
\eeq
In this matrix representation Poincar\'e transformations continuously
connected to the identity are represented by
\beq
X \to X' = \Lambda X \Lambda^{\dagger} + A \qquad \Lambda \in SL(2 ,\mathbb{C}) \qquad A=A^{\dagger} .
\label{n.7}
\eeq
The subgroup of the Poincar\'e group that leaves $x^+=0$ 
invariant consists of pairs of matrices $(\Lambda,A)$ in (\ref{n.7})
of the form
\beq
\Lambda =
\left (
\begin{array}{cc}
a & 0\\
c & 1/a \\
\end{array}
\right )
\qquad 
A =
\left (
\begin{array}{cc}
0 & b_{\perp}^*\\
b_{\perp}  & b^- \\
\end{array}
\right )
\label{n.8}
\eeq
where $a$,$c$ and $b_{\perp}$ are complex and $b^-$ is real.  This is a
seven- parameter group.
The $SL(2,\mathbb{C})$ matrices with real $a$ represent light-front preserving
boosts.  They can be parameterized by the light-front components of the
four velocity $v=p/m$:
\beq
\Lambda_f (p/m) := 
\left (
\begin{array}{cc} 
\sqrt{p^+/m}  &0   \\
{p_{\perp}/m \over
 \sqrt{p^+/m}}& 1/\sqrt{p^+/m}  \\
\end{array} 
\right ) =
\left (
\begin{array}{cc} 
\sqrt{v^+}  &0   \\
v_{\perp} /
 \sqrt{v^+}& 1/\sqrt{v^+}  \\
\end{array} 
\right ).
\label{n.9}
\eeq
These lower triangular matrices form a subgroup.  The inverse light-front
boost is given by
\beq
\Lambda_f^{-1} (p/m) :=
\left (
\begin{array}{cc} 
1/\sqrt{p^+/m}  &0   \\
-{p_{\perp}/m \over
 \sqrt{p^+/m}}& \sqrt{p^+/m}  \\
\end{array} 
\right )=
\left (
\begin{array}{cc} 
1/ \sqrt{v^+}  &0   \\
-v_{\perp} /
\sqrt{v^+}& \sqrt{v^+}  \\
\end{array} 
\right )
\label{n.10}
\eeq
while the adjoint and the inverse adjoint of these matrices are
\beq
\Lambda_f^{\dagger}  (p/m) :=
\left (
\begin{array}{cc} 
\sqrt{p^+/m}  &{p_{\perp}^*/m \over
 \sqrt{p^+/m}}   \\
0& 1/\sqrt{p^+/m}  \\
\end{array} 
\right )=
\left (
\begin{array}{cc} 
\sqrt{v^+}  &v_{\perp}^* /\sqrt{v^+}   \\
0& 1/\sqrt{v^+}  \\
\end{array} 
\right )
\label{n.11}
\eeq 
\beq
(({\Lambda}_f)^{\dagger})^{-1} (p/m) := 
\left (
\begin{array}{cc} 
1/\sqrt{p^+/m}  &-{p_{\perp}^*/m \over
 \sqrt{p^+/m}}   \\
0& \sqrt{p^+/m}  \\
\end{array} 
\right )=
\left (
\begin{array}{cc} 
1/\sqrt{v^+}  &-v_{\perp}^* /\sqrt{v^+}   \\
0& \sqrt{v^+}  \\
\end{array} 
\right ).
\label{n.12}
\eeq
General Poincar\'e transformations are generated by 10 independent
one-parameter subgroups.  Seven of the one-parameter groups leave the
light front invariant.  The remaining three one-parameter groups map
points on the light front to points off of the light front.  These are
called kinematic and dynamical transformations respectively.  The
kinematic one-parameter groups in the $2 \times 2$ matrix
representation and the corresponding unitary representations of these
groups are related by
\beq
\Lambda (\lambda) =
\left (
\begin{array}{cc}
1 & 0 \\
\lambda & 1 \\
\end{array}
\right )
\qquad
U(\Lambda (\lambda)) = e^{i E^1 \lambda}
\qquad
\Lambda (\lambda) =
\left (
\begin{array}{cc}
1 & 0 \\
i \lambda & 1 \\
\end{array}
\right )
\qquad
U(\Lambda (\lambda)) = e^{i E^2 \lambda}
\label{n.12.1}
\eeq
\beq
\Lambda (\lambda) =
\left (
\begin{array}{cc}
e^{\lambda /2} & 0 \\
0 & e^{-\lambda /2} \\
\end{array}
\right )
\qquad
U(\Lambda (\lambda)) = e^{i K^3 \lambda}
\qquad
\Lambda (\lambda) =
\left (
\begin{array}{cc}
e^{i\lambda /2} & 0 \\
0 & e^{-i\lambda /2} \\
\end{array}
\right )
\qquad
U(\Lambda (\lambda)) = e^{i J^3 \lambda}
\label{n.13}
\eeq
\beq
A (\lambda) =
\left (
\begin{array}{cc}
0 & \lambda \\
\lambda &  0\\
\end{array}
\right )
\qquad
U(\Lambda (\lambda)) = e^{i P^1 \lambda }.
\qquad 
A (\lambda) =
\left (
\begin{array}{cc}
0 & -i\lambda \\
i\lambda &  0\\
\end{array}
\right )
\qquad
U(\Lambda (\lambda)) = e^{iP^2 \lambda }.
\label{n.14a}
\eeq
\beq
A (\lambda) =
\left (
\begin{array}{cc}
0 & 0 \\
0 &  \lambda\\
\end{array}
\right )
\qquad
U(\Lambda (\lambda)) = e^{-{i\over 2} P^+ \lambda }.
\label{n.14}
\eeq

The corresponding dynamical transformations are
\beq
\Lambda (\lambda) =
\left (
\begin{array}{cc}
1 & \lambda \\
0 & 1 \\
\end{array}
\right )
\qquad
U(\Lambda (\lambda)) = e^{i F^1 \lambda}
\qquad
\Lambda (\lambda) =
\left (
\begin{array}{cc}
1 & - i \lambda  \\
0 & 1 \\
\end{array}
\right )
\qquad
U(\Lambda (\lambda)) = e^{i F^2 \lambda}
\label{n.15}
\eeq
\beq
A (\lambda) =
\left (
\begin{array}{cc}
\lambda & 0 \\
0 &  0\\
\end{array}
\right )
\qquad 
U(\Lambda (\lambda)) = e^{-{i\over 2} P^- \lambda}.
\label{n.16}
\eeq
Relations (\ref{n.12.1}-\ref{n.14}) define
the infinitesimal generators 
\beq
\{ P^+, P^1,P^2,E^1,E^2,K^3,J^3 \}
\label{n.17}
\eeq
of the kinematic transformations,
while (\ref{n.15}-\ref{n.16}) define the infinitesimal generators
\beq
\{ P^-,F^1,F^2\}
\label{n.18}
\eeq
of the dynamical transformations.
With these definitions the light-front Poincar\'e generators are related to
components of the angular momentum tensor
\beq
J^{\mu\nu}
=
\left (
\begin{array}{cccc}
0 &-K^1& -K^2& -K^3\\
K^1 & 0 & J^3 & -J^2\\
K^2 & -J^3 & 0 &J^1\\
K^3 & J^2 & -J^1 &0 \\
\end{array}
\right )
\label{n.19}
\eeq  
by
\beq
E^1= K^1-J^2 \qquad
E^2= K^2+J^1 \qquad
F^1= K^1+J^2 \qquad
F^2= K^2-J^1 .
\label{n.20}
\eeq
The inverse relations are
\beq
K^1={1\over 2} (E^1+F^1) \qquad
K^2={1\over 2} (E^2+F^2) \qquad
J^1={1\over 2} (E^2-F^2) \qquad
J^2={1\over 2} (F^1-E^1) .
\label{n.21}
\eeq
$F^1$ and $F^2$ could be
replaced by $J^1$ and $J^2$ as dynamical generators.

The evolution of a state or operator with initial data on the light front
is determined by the light-front Schr\"odinger equation 
\beq
i {d \vert \psi (x^+) \rangle \over d x^+} =
{1 \over 2} P^- \vert \psi (x^+) \rangle
\label{n.22}
\eeq
or the light-front Heisenberg equations of motion 
\beq
{d O(x^+) \over dx^+} = {i \over 2} [P^-,O(x^+)].
\label{n.23}
\eeq
When $P^-$ is a self-adjoint operator the dynamics is well-defined
and given by the unitary one-parameter group (\ref{n.16}).

The Poincar\'e Lie algebra has two polynomial invariants.
The mass squared is
\beq
M^2 = P^+ P^- - \mathbf{P}_{\perp}^2
\label{n.24}
\eeq
which gives the light-front dispersion relation
\beq
P^- = {M^2 +\mathbf{P}^2 \over P^+} .
\label{n.25}
\eeq
The other invariant is the inner product of the
Pauli-Lubanski vector,
\beq
W^{\mu} = {1 \over 2} \epsilon^{\mu \nu \alpha \beta}
P_{\nu}J_{\alpha \beta},
\label{n.26}
\eeq
with itself
\beq
W^2 = W^{\mu}W_{\mu} =  M^2 s^2 .
\label{n.27}
\eeq
The Pauli-Lubanski vector has components
\beq
W^0 = \mathbf{P}\cdot \mathbf{J} \qquad
\mathbf{W} = H\mathbf{J} + \mathbf{P}\times \mathbf{K} 
\label{n.28}
\eeq
or expressed in terms of the light-front Poincar\'e generators
\beq
W^+ = P^+ \mathbf{J} \cdot \hat{\mathbf{z}} + (\mathbf{P} \times \mathbf{E})\cdot \hat{\mathbf{z}}
\label{n.29}
\eeq
\beq
\mathbf{W}_{\perp}= {1 \over 2} (
P^+ \hat{\mathbf{z}} \times \mathbf{F}
-P^- \hat{\mathbf{z}} \times \mathbf{E}) -
(\hat{\mathbf{z}} \cdot \mathbf{K}) \hat{\mathbf{z}} \times \mathbf{P}
\label{n.30}
\eeq
\beq
W^- = P^- \mathbf{J} \cdot \hat{\mathbf{z}} - (\mathbf{P} \times \mathbf{F})\cdot \hat{\mathbf{z}}.
\label{n.31}
\eeq
In order to compare the spins of particles in different frames, it is
useful to transform both particles to their rest frame
using an arbitrary but fixed set of Lorentz transformations parameterized
by the four velocity of the particle.
The light-front spin is the angular momentum measured in the particle
or system rest frame when the particles or system are transformed
to the rest frame with the inverse light-front preserving boosts (\ref{n.10})
\beq
\mathbf{s} \cdot \hat{\mathbf{z}} = 
\mathbf{J} \cdot \hat{\mathbf{z}} - {(\mathbf{E} \times \mathbf{P})\cdot
\hat{\mathbf{z}} \over P^+} = {W^+ \over P^+} 
\label{n.33}
\eeq
\beq
\mathbf{s}_{\perp}= (\mathbf{W}_{\perp} - \mathbf{P}_{\perp}W^+/P^+)/M . 
\label{n.34}
\eeq
The components of the light-front spin can also be expressed
directly in terms of $J^{\mu \nu}$
\beq
s^i  = {1 \over 2} \epsilon^{ijk} (\Lambda^{-1}_{lf})^j{}_{\mu}(P/M)
(\Lambda^{-1}_{lf})^k{}_{\nu}(P/M)J^{\mu \nu}
\label{n.35}
\eeq
where in (\ref{n.35}) the $P/M$ in the Lorentz boosts are operators.

\section{Fields}

Light-front free fields can be constructed from canonical free fields
by changing variables $\mathbf{p} \to \tilde{\mathbf{p}}$, where
$\tilde{\mathbf{p}} := (p^+,p^1,p^2)$ are the components of the
light front-momentum conjugate
to $\tilde{\mathbf{x}}$.
The Fourier representation of a free scalar
field of mass $m$ and its conjugate momentum operator are
\beq
\phi (x) = {1 \over (2 \pi)^{3/2}}
\int {d\mathbf{p} \over \sqrt{2\omega_m(\mathbf{p})}}
\left  (e^{i p\cdot x} a(\mathbf{p}) +
e^{-i p\cdot x} a^{\dagger}(\mathbf{p}) \right) 
\label{f.1}
\eeq
\beq
\pi (x) = -{i \over (2 \pi)^{3/2}}
\int d\mathbf{p}\sqrt{{\omega_m(\mathbf{p})\over 2}}
\left  (e^{i p\cdot x} a(\mathbf{p}) -
e^{-i p\cdot x} a^{\dagger}(\mathbf{p}) \right)
\label{f.2}
\eeq
where $\omega_m(\mathbf{p}):= \sqrt{m^2 +\mathbf{p}^2}$ is the energy
of a particle of mass $m$,  $\mathbf{p}$ is its three-momentum and $x \cdot p := - \omega_m(\mathbf{p})x^0 + \mathbf{p}\cdot \mathbf{x}$.

Changing variables from the three momentum, $\mathbf{p}$, to the light-front
components, $\tilde{\mathbf{p}}= (p^+,p^1,p^2)$,
of the four momentum gives the light-front Fourier representation
of $\phi(x)$:
\beq
\phi (x) = {1 \over (2 \pi)^{3/2}}
\int {dp^+ \theta (p^+) \over \sqrt{2p^+}}d\mathbf{p}_{\perp}
\left  (e^{i p\cdot x} \tilde{a}(\tilde{\mathbf{p}}) +
e^{-i p\cdot x} \tilde{a}^{\dagger}(\tilde{\mathbf{p}}) \right)
\label{f.3}
\eeq
where
\beq
\vert { \partial (p^1, p^2, p^2)\over  \partial (p^+, p^1, p^2)}\vert
= {\omega_m (\mathbf{p})\over p^+}
\qquad
p \cdot x = -{1 \over 2}({\mathbf{p}^2_{\perp}+m^2 \over p^+}x^+
+ p^+x^-) + \mathbf{p}_{\perp}\cdot \mathbf{x}_{\perp}
\label{f.4}
\eeq
and
\beq
\tilde{a}(\tilde{\mathbf{p}}) := 
\tilde{a} (p^+,\mathbf{p}_{\perp}) =  a(\mathbf{p})
\sqrt{{\omega_m (\mathbf{p})\over p^+}}.
\label{f.5}
\eeq
It follows from 
\beq
[a (\mathbf{p}),a^{\dagger} (\mathbf{p}')]= \delta (\mathbf{p}-\mathbf{p}')
\label{f.6}
\eeq
and (\ref{f.4}) and (\ref{f.5}) that
\beq
[a (\tilde{\mathbf{p}}),a^{\dagger} (\tilde{\mathbf{p}}')]=
\delta (\tilde{\mathbf{p}}-\tilde{\mathbf{p}}').
\eeq
The spectral conditions 
\beq
P^\pm = H \pm P^3 = \sqrt{M^2 +\mathbf{P}^2} \pm P^3 \geq 0   
\label{f.7}
\eeq
\beq
P^- = {M^2 +\mathbf{P}^2 \over P^+} \geq 0
\label{f.8}
\eeq
imply that it is possible to independently construct both $a
(\tilde{\mathbf{p}})$ and $a^{\dagger} (\tilde{\mathbf{p}})$ from the
field $\phi (x^+=0,\tilde{\mathbf{x}})$ restricted to the light front.
This can be done by computing the partial Fourier transform of the field on
the light front:
\beq
\phi (x^+=0,p^+,\mathbf{p}_{\perp})=
{1 \over (2\pi)^{3/2}}
\int e^{i p^+ x^-/2 -i \mathbf{p}_{\perp}\cdot \mathbf{x}_{\perp}}
\phi (x^+=0,x^-,\mathbf{x}_{\perp}){d\mathbf{x}_{\perp}dx^- \over 2}.
\label{f.9}
\eeq
The creation and annihilation operators can be read off of this
expression
\beq
\tilde{a}(\tilde{\mathbf{p}}) = \sqrt{{p^+\over 2}}\theta (p^+)
\phi (x^+=0,p^+,\mathbf{p}_{\perp})
\label{f.10}
\eeq
\beq
\tilde{a}^{\dagger} (\tilde{\mathbf{p}}) = \sqrt{p^+\over 2}\theta (p^+)
\phi (x^+=0,-p^+,\mathbf{p}_{\perp}) .
\label{f.11}
\eeq
Both operators are constructed directly from the field restricted to the
light front without constructing a generalized
momentum operator.  This means that $\phi (x)$ restricted to the light
front defines an irreducible set of operators.  It follows that any
operator $O$ on the Fock space that
commutes with $\phi (x^+=0,\tilde{\mathbf{x}})$ at
all points on the light front must be a constant multiple of the
identity:
\beq
[\phi (x^+=0,\tilde{\mathbf{x}}),O]=0 \to O= cI .
\label{f.12}
\eeq
An important observation is that the only place where the mass of the
field appears is in the expression for the coefficient of $x^+$.  When
the field is restricted to the light front, $x^ +\to 0$, all information
about the mass (and dynamics) disappears.

This is in contrast to the canonical case because the canonical
transformation that relates free canonical fields and their
generalized momenta with different masses cannot be realized by a
unitary transformation \cite{Haag:1955ev}.  When these fields are
restricted to the light front they become unitarily equivalent
\cite{Schlieder:1972qr}.  This is because dynamical information that
distinguishes the different representations is lost as a result of the
restriction.

Since the fields restricted to the light front are irreducible, 
the canonical commutation relations are replaced by the commutator
of the fields at different points on the light front
\beq
[\phi (x^+=0, \tilde{\mathbf{x}}),\phi (y^+=0, \tilde{\mathbf{y}})]=
{i \over 2\pi} \int {dp^+ \theta (p^+)\over p^+}
{e^{-{i\over 2} p^+ (x^--y^-)} - e^{{i\over 2} p^+ (x^--y^-)}\over 2i}
\delta (\mathbf{x}_{\perp}-\mathbf{y}_{\perp}) =
\label{f.13}
\eeq
\beq
-{i \over 2\pi} \int {dp^+ \theta (p^+)\over p^+}
\sin ({1\over 2} p^+ (x^--y^-)) \delta (\mathbf{x}_{\perp}-\mathbf{y}_{\perp})
= -{i \over 4}\epsilon (x^--y^-)\delta (\mathbf{x}_{\perp}-\mathbf{y}_{\perp}).
\label{f.14}
\eeq
Note that while the $x^-$ derivative gives
\beq
{\partial \over \partial x^-} [\phi (x^+=0, \tilde{\mathbf{x}}),\phi (y^+=0, \tilde{\mathbf{y}})] =
-{i \over 2}\delta (x^--y^-)\delta (\mathbf{x}_{\perp}-\mathbf{y}_{\perp}),
\label{f.15}
\eeq
$\partial_- \phi (x)$  is not the canonical momentum.

Interactions that preserve the light-front kinematic symmetry must
commute with the kinematic subgroup.  In particular they must be
invariant with respect to translations in the $x^-$ direction.  This
means that the interactions must commute with $P^+$, which is a
kinematic operator.  Since, $P^+= \sum_i P^+_i$, is kinematic, the
vacuum of the field theory is invariant with respect to these
translations, independent of interactions.  This requires that
\beq
[P^+,V]=0 \qquad P^+ \vert 0 \rangle =0
\label{f.15}
\eeq
which implies
\beq
P^+ V \vert 0 \rangle = V P^+ \vert 0 \rangle =0
\label{f.15}
\eeq
where $\vert 0 \rangle$ is the free-field Fock vacuum.
This means that $V \vert 0 \rangle$ is an eigenstate of $P^+$ with
eigenvalue $0$. Inserting a compete set of intermediate states between
$V^{\dagger}$ and $V$ in $\langle 0 \vert V^{\dagger} V \vert 0
\rangle$, the absolutely continuous spectrum of $p^+_i$ cannot
contribute to the sum over intermediate states because $p_i^+=0$ is a
set of measure 0.  This means that
\beq
V \vert 0 \rangle = \vert 0 \rangle \langle 0 \vert V \vert 0 \rangle
\label{f.15.1}
\eeq
or interactions that preserve the
kinematic symmetry leave the free-field Fock vacuum unchanged.

The observation that the interaction leaves the vacuum invariant
implies that it is an operator on the free-field Fock space.  The
irreducibility of the light front-Fock algebra means that the
interaction can be expressed in terms of fields in this algebra. 
The Poincar\'e generators, defined by integrating the $+$
components of the Noether currents that come from Poincar\'e
invariance of the action over the light front, are also linear in this
interaction.  This means that it should be possible to solve for
the relativistic dynamics of the field on the light-front Fock
space.

A more careful analysis shows that the interaction, while formally
leaving the light front invariant, has singularities at $p^+=0$, so
the formal expressions for the interaction-dependent generators are
not well-defined self-adjoint operators on the free field Fock space.
This is because the interaction contains products of operator-valued
distributions which are not defined.  Discussions of the
non-triviality of the light-front vacuum and the associated ``zero-mode''
problem, which is the subject of many papers, can be found in
\cite{chang:1969}\cite{yan}
\cite{nakanishi1}\cite{nakanishi2}
\cite{lenz}\cite{werner}\cite{collins}\cite{marc} and the references
cited therein.

The expressions for the Poincar\'e generators are defined on the
free-field Fock space if infrared and ultraviolet cutoffs are
introduced, but the cutoffs break the Poincar\'e symmetry.  The
non-trivial problem is how to remove the cutoffs in a manner that
recovers the Poincar\'e symmetry.

While the solution of this last problem is equivalent to the unsolved
problem of giving a non-perturbative definition of the theory,
cutoff theories should  lead to good approximations for observables
on scales where the cutoffs are not expected to be important.

\section{formal light-front field dynamics}

The Lagrangian density for a scalar field theory is 
\beq
{\cal L}(\phi(x))
= -{1 \over 2} \eta^{\mu \nu}
\partial_{\mu}\phi(x) \partial_{\nu} \phi (x) 
-{1 \over 2} m^2 \phi(x)^2 -  V(\phi(x)) 
\label{fd.1}
\eeq
where $\eta^{\mu\nu}$ is the metric tensor with signature $(-,+,+,+)$.
The action is
\beq
A[V,\phi] = \int_{V} d^4 x {\cal L}(\phi (x)).
\label{fd.2}
\eeq
Variations of the field that leave the action stationary satisfy the
field equation:
\beq
{\partial^2 \phi (x) \over \partial (x^0)^2} - \pmb{\nabla}^2\phi (x)
+ m^2 \phi (x) + {\partial V(\phi) \over \partial \phi (x)}=0.
\label{fd.3}
\eeq
Changing to light-front variables the partial derivatives become
\beq
\partial_0 := {\partial \over \partial x^0} =
{\partial x^+\over \partial x^0} {\partial \over \partial x^+} +
{\partial x^-\over \partial x^0} {\partial \over \partial x^-} 
={\partial \over \partial x^+} +
{\partial \over \partial x^-}= \partial_+ + \partial_- 
\label{fd.4}
\eeq
\beq
\partial_3 := {\partial \over \partial x^3} =
{\partial x^+\over \partial x^3} {\partial \over \partial x^+} +
{\partial x^-\over \partial x^3} {\partial \over \partial x^-} 
=
{\partial \over \partial x^+} -
{\partial \over \partial x^-}= \partial_+ - \partial_-. 
\label{fd.5}
\eeq
Squaring and subtracting gives
\beq
{\partial^2 \over \partial (x^0)^2} -
{\partial^2 \over \partial (x^3)^2} =
4{\partial \over \partial x^+}
{\partial \over \partial x^-}. 
\label{fd.6}
\eeq
It follows that the Lagrangian density (\ref{fd.1})
and the field equation
in light-front variables
have the forms
\beq
{\cal L}(\phi(x))=
2 \partial_-  \phi(x) \partial_+ \phi(x)
-{1 \over 2} \pmb{\nabla}_{\perp}\phi(x)\cdot\pmb{\nabla}_{\perp}\phi(x)
- {1 \over 2} m^2 \phi(x)^2 - V(\phi(x))
\label{fd.7}
\eeq
and
\beq
4\partial_+ 
\partial_-  \phi (x)
- \pmb{\nabla}_{\perp} ^2\phi (x)
+ m^2 \phi (x) + {\partial V(\phi) \over \partial \phi (x)} =0.
\label{fd.8}
\eeq
Invariance of the action under infinitesimal
changes in the fields and coordinates
\beq
\phi (x) \to \phi' (x') = \phi(x) + \delta \phi (x)
\qquad
x^{\mu} \to x^{\prime \mu} + \delta x^{\mu}(x),
\label{fd.9}
\eeq
along with the field equation, leads to the conserved Noether currents 
\beq
\partial_{\mu} J^{\mu} (x) =0 
\label{fd.10}
\eeq
where the Noether current is
\beq
J^{\mu} (x) = {\cal L}\eta^{\mu \nu} \delta x_{\nu} 
+ {\partial {\cal L}(\phi) \over \partial(\partial_{\mu}\phi)}
(\delta \phi (x) - \partial_{\nu}\delta x^{\nu}).
\label{fd.11}
\eeq
The Noether currents associated with translational and Lorentz
invariance of the action are the energy momentum, $T^{\mu \nu}$, and
angular momentum $M^{\mu \alpha \beta}$ tensors
\beq
\partial_{\mu}  T^{\mu \nu} = 0
\qquad
\partial_{\mu}  M^{\mu \alpha \beta} = 0
\label{fd.12}
\eeq
where for the Lagrangian density (\ref{fd.7}):
\beq
T^{\mu \nu} =
\eta^{\mu \nu} {\cal L}(\phi(x))  +
\partial^{\mu} \phi (x) \partial^{\nu} \phi (x)
\label{fd.13}
\eeq
\beq
M^{\mu \alpha \beta} = 
T^{\mu\alpha }x^{\beta} - T^{\mu\beta }x^{\alpha}  .
\label{fd.14}
\eeq   
Integrating the $+$ component of the conserved current over the
light front, assuming that the
fields vanish on the boundary of the light front,  give
the light-front conserved (independent of $x^+$) charges
\beq
{d \over dx^+} P^{\mu} =0 \qquad {d \over dx^+} J^{\alpha\beta} =0
\label{fd.15}
\eeq
where 
\beq
P^{\mu} := \int {d \mathbf{x}_{\perp}dx^- \over 2}
T^{+\mu} = \int {d \mathbf{x}_{\perp}dx^- \over 2}
(T^{0\mu} + T^{3\mu})
\label{fd.16}
\eeq  
and
\beq
J^{\alpha \beta} :=  
\int {d \mathbf{x}_{\perp}dx^- \over 2} M^{+\alpha\beta}
=\int {d \mathbf{x}_{\perp}dx^- \over 2}
((T^{0\alpha}+ T^{3\alpha})x^{\beta} - (T^{0\beta}+ T^{3\beta})x^{\alpha}) .
\label{fd.17}
\eeq
These are the conserved four momentum and angular-momentum tensors.
They are independent of $x^+$ and thus can be expressed in terms of
fields and derivatives of fields restricted to the light front.

In order to construct the Poincar\'e generators the first step is to express the
$+$ component of the energy-momentum tensor and angular-momentum tensors
in terms of fields on the light front:
\beq
T^{++} 
=
4 \partial_-  \phi(x) \partial_-  \phi(x)
\label{fd.18}
\eeq
\beq
T^{+i} = -2 \partial_-  \phi(x) \partial_i \phi(x) 
\label{fd.19}
\eeq
\beq
T^{+-} =  \pmb{\nabla}_{\perp}\phi(x)\cdot\pmb{\nabla}_{\perp}\phi(x)
+ m^2 \phi^2(x) + 2 V(\phi(x))
\label{fd.20}
\eeq
\beq
M^{++-}=
4 \partial_-\phi(x) \partial_-\phi(x)x^- -
( \pmb{\nabla}_{\perp}\phi(x)\cdot\pmb{\nabla}_{\perp}\phi(x)
+ m^2 \phi^2(x) + 2 V(\phi(x))x^+
\label{fd.21}
\eeq
\beq
M^{++i}=
4 \partial_-\phi(x) \partial_-\phi(x)x^i +
2 \partial_-\phi(x) \partial_i\phi(x)x^+
\label{fd.22}
\eeq
\beq
M^{+-i}=
( \pmb{\nabla}_{\perp}\phi(x)\cdot\pmb{\nabla}_{\perp}\phi(x)
+ m^2 \phi^2(x) + 2 V(\phi(x))x^i
+ 2 \partial_-\phi(x) \partial_i\phi(x)x^-
\label{fd.23}
\eeq
\beq
M^{+ij}=
-2 \partial_-\phi(x) \partial_i\phi(x)x^j +
2 \partial_-\phi(x) \partial_j\phi(x)x^i .
\label{fd.24}
\eeq
The Poincar\'e generators are constructed by integrating these operators over
the light front
\beq
P^{+} = 4 \int {dx^- d^2x_{\perp} \over 2}  \partial_-  \phi(x) \partial_-  \phi(x)
\label{fd.25}
\eeq
\beq
P^{i} = -2 \int {dx^- d^2x_{\perp} \over 2}\partial_-  \phi(x) \partial_i \phi(x) 
\label{fd.26}
\eeq
\beq
P^{-} = \int {dx^- d^2x_{\perp} \over 2}\left ( \pmb{\nabla}_{\perp}\phi(x)\cdot\pmb{\nabla}_{\perp}\phi(x)
+ m^2 \phi^2(x) + 2 V(\phi(x)\right )
\label{fd.27}
\eeq
\beq
J^{+-}=
\int {dx^- d^2x_{\perp} \over 2}
\left ( 
4 \partial_-\phi(x) \partial_-\phi(x)x^- -
( \pmb{\nabla}_{\perp}\phi(x)\cdot\pmb{\nabla}_{\perp}\phi(x)
+ m^2 \phi^2(x) + 2 V(\phi(x))x^+
\right )
\label{fd.28}
\eeq
\beq
J^{+i}=\int {dx^- d^2x_{\perp} \over 2}
\left (
4 \partial_-\phi(x) \partial_-\phi(x)x^i +
2 \partial_-\phi(x) \partial_i\phi(x)x^+
\right )
\label{fd.29}
\eeq
\beq
J^{-i}=\int {dx^- d^2x_{\perp} \over 2}
\left (
( \pmb{\nabla}_{\perp}\phi(x)\cdot\pmb{\nabla}_{\perp}\phi(x)
+ m^2 \phi^2(x) + 2 V(\phi(x))x^i
+ 2 \partial_-\phi(x) \partial_i\phi(x)x^-
\right )
\label{fd.30}
\eeq
\beq
J^{ij}= \int {dx^- d^2x_{\perp} \over 2}
\left (
-2 \partial_-\phi(x) \partial_i\phi(x)x^j +
2 \partial_-\phi(x) \partial_j\phi(x)x^i
\right ).
\label{fd.31}
\eeq
For free fields these operators
can be expressed in terms of the light-front creation and annihilation
operators (\ref{f.10}-\ref{f.11}) using the identities
\beq
\int {dx^- d^2x_{\perp} \over 2} :\phi (x) \phi (x):=
\int {\theta (p^+) dp^+ d^2 p_{\perp} \over p^+}
 \tilde{a}^{\dagger} (\tilde{p}) \tilde{a} (\tilde{p})
\label{fd.32}
 \eeq
\beq
\int {dx^- d^2x_{\perp} \over 2} :\partial_-\phi (x) \partial_- \phi (x):=
{1 \over 4} \int \theta (p^+) dp^+ d^2 p_{\perp}
\tilde{a}^{\dagger} (\tilde{p}) p^+\tilde{a} (\tilde{p})
\label{fd.33}
\eeq
\beq
\int {dx^- d^2x_{\perp} \over 2} :\partial_-\phi (x) \partial_i \phi (x):=
-{1 \over 2} \int \theta (p^+) dp^+ d^2 p_{\perp}
\tilde{a}^{\dagger} (\tilde{p}) p^i\tilde{a} (\tilde{p})
\label{fd.34}
\eeq
\beq
\int {dx^- d^2x_{\perp} \over 2} :\partial_i\phi (x) \partial_i\phi (x):=
\int {\theta (p^+) dp^+ d^2 p_{\perp} \over p^+}
\tilde{a}^{\dagger} (\tilde{p}) (p^i)^2\tilde{a} (\tilde{p}).
\label{fd.35}
\eeq
Using (\ref{fd.32}-\ref{fd.35}) in (\ref{fd.25}-\ref{fd.31}) gives the following expressions
for the Poincar\'e generators for a free field in terms of the
light-front creation and annihilation operators
\beq
P^+= \int dp^+ d^2p_{\perp} \theta (p^+) 
\tilde{a}^{\dagger} (\tilde{p}) p^+ \tilde{a} (\tilde{p})
\label{fd.36}
\eeq
\beq
P^i=\int dp^+ d^2p_{\perp} \theta (p^+) 
\tilde{a}^{\dagger} (\tilde{p}) p^i\tilde{a} (\tilde{p})
\label{fd.37}
\eeq
\beq
P^-=\int dp^+ d^2p_{\perp} \theta (p^+) 
\tilde{a}^{\dagger} (\tilde{p}) {\mathbf{p}_{\perp}^2 +m^2 \over p^+}
\tilde{a} (\tilde{p})
\label{fd.38}
\eeq
\beq
J^{+-}=
\int dp^+ d^2p_{\perp} \theta (p^+) 
\tilde{a}^{\dagger} (\tilde{p})(
p^+ (-2 i {\partial \over \partial p^+}) -
x^+ {\mathbf{p}_{\perp}^2 + m^2 \over p^+} 
)\tilde{a} (\tilde{p}) 
\label{fd.39}
\eeq
\beq
J^{+i}=
\int dp^+ d^2p_{\perp} \theta (p^+) 
\tilde{a}^{\dagger} (\tilde{p})(
(p^+ (i {\partial \over \partial p^i}) -
p^i x^+  
)\tilde{a} (\tilde{p})
\label{fd.40}
\eeq
\beq
J^{-i}=
\int dp^+ d^2p_{\perp} \theta (p^+) 
\tilde{a}^{\dagger} (\tilde{p})(
{\mathbf{p}_{\perp}^2 + m^2 \over p^+}(i {\partial \over \partial p^i})
- 2 p^i (-i {\partial \over \partial p^+}) 
)\tilde{a} (\tilde{p})
\label{fd.41}
\eeq
\beq
J^{ij}=
\int dp^+ d^2p_{\perp} \theta (p^+) 
\tilde{a}^{\dagger} (\tilde{p})(
p^j (-i {\partial \over \partial p^i}-
p^i (-i {\partial \over \partial p^j})
)\tilde{a} (\tilde{p}).
\label{fd.42}
\eeq
Since these are independent of $x^+$, the expressions
with an explicit $x^+$ dependence can be evaluated at $x^+=0$.
These expressions lead to the following identifications
\beq
J^{+-} = -2 K^3
\qquad
J^{+1}=K^1-J^2 = E^1 \qquad
J^{+2}=K^2+J^1 = E^2
\label{fd.43}
\eeq
\beq
J^{-1}=K^1+J^2 = F^1 \qquad
J^{-2}=K^2-J^1 = F^2 .
\label{fd.44}
\eeq

\section{wavelet basis}

In this section the multi-resolution basis that is used to represent the
irreducible algebra of fields on the light front is introduced. 
Wavelets provide a natural means for exactly decomposing a field into
independent discrete degrees of freedom labeled by volume and
resolution.  In this representation there are natural truncations that
eliminate degrees of freedom associated with volumes and resolutions
that are expected to be unimportant in modeling a given reaction.

While there are many different types of wavelets, this
application uses Daubechies \cite{daubechies:1988}\cite{daubechies}
$L=3$ wavelets.  These are used to generate an orthonormal basis of
functions with the following desirable properties: (1) all of the
basis functions have compact support (2) there are an infinite number
of basis functions with compact support inside of any open set (3) the
basis functions have one continuous derivative (4) polynomials of
degree 2 can be point-wise represented by locally-finite linear
combinations of these basis functions.

In what follows these basis functions will be used to decompose fields
restricted to a light front into an infinite linear combination of
discrete operators with arbitrarily fine resolutions.  The advantage
of the light-front representation is that the resulting discrete algebra
is irreducible and the vacuum remains trivial.

For Lagrangians that are polynomials in the fields, in the wavelet
representation all of the Poincar\'e generators can be formally
expressed as polynomials in the discrete fields on the light front
with coefficients that can be computed analytically.  While the
polynomials are finite degree, there are an infinite number of
discrete field operators.

The construction of the wavelet basis starts with the fixed-point solution of
the renormalization group equation
\beq
s(x) = \sum_{l=0}^{2L-1}h_l D T^l s(x)
\label{wav.1}
\eeq
where
\beq
Df(x):= \sqrt{2}f(2x) \qquad \mbox{and} \qquad Tf(x) := f(x-1)
\label{wav.2}
\eeq
are unitary scale transformations and translations.
The fixed point, $s(x)$,
is a
linear combination of a weighted sum of translates of itself on a
smaller scale by a factor of 2.
The weights $h_l$ are
constant coefficients chosen so $s(x)$ satisfies
\beq
\int T^m s(x) T^n s(x) = \delta_{mn} \qquad \mbox{and} \qquad
x^k = \sum_n c_n^k T^n s(x) \qquad k<L \qquad \mbox{point-wise}.
\label{wav.3}
\eeq
There are different weights $h_l$ for different values of $L$.
The $L=3$ weights are the algebraic numbers in table 1.
Solving (\ref{wav.1}) is analogous to finding a fixed point of a block spin
transformation, except the averaging over blocks is replaced by a
weighted average.

\begin{table}[t]
\caption{Scaling Coefficients for Daubechies L=3 Wavelets}
\begin{tabular}{|l|l|}
\hline				      		      
$h_0$ & $(1+\sqrt{10}+\sqrt{5+2\sqrt{10}}\,)/16\sqrt{2}$ \\
$h_1$ & $(5+\sqrt{10}+3\sqrt{5+2\sqrt{10}}\,)/16\sqrt{2}$ \\
$h_2$ & $(10-2\sqrt{10}+2\sqrt{5+2\sqrt{10}}\,)/16\sqrt{2}$ \\
$h_3$ & $ (10-2\sqrt{10}-2\sqrt{5+2\sqrt{10}}\,)/16\sqrt{2} $ \\
$h_4$ & $(5+\sqrt{10}-3\sqrt{5+2\sqrt{10}}\,)/16\sqrt{2}$ \\
$h_5$ & $(1+\sqrt{10}-\sqrt{5+2\sqrt{10}}\,)/16\sqrt{2}$ \\
\hline
\end{tabular}
\label{coef}
\end{table}

The solution of the renormalization group
equation (\ref{wav.1}) is a fractal valued function that
has compact support for $x \in [0,2L-1]$.  For $L=3$ the
solution has one continuous derivative with support on the interval
$[0,5]$.  
Since the scale can be changed by a general unitary scale transformation, a
scale is fixed by the convention
\beq
\int s(x) dx =1 .
\label{wav.4}
\eeq

Because $s(x)$ is fractal valued it cannot be represented in terms of
elementary functions, however it can be exactly calculated at all
dyadic rationals using the renormalization group equation
(\ref{wav.1}).  It can also be approximated by iterating the
renormalization group equation starting with a seed function
satisfying (\ref{wav.4}).  The evaluation of $s(x)$ is not necessary
because most of the integrals that are needed in field theory
applications can be evaluated exactly using the renormalization group
equation.  The integrals can be expressed in terms of solutions of
finite linear systems of equations involving the numerical weights
$h_l$ in table 1.

The next step in constructing the wavelet basis is to construct subspaces of
$L^2(\mathbb{R})$ with different resolutions defined by
\beq
{\cal S}^k := \{ f(x) \vert f(x) = \sum_n c_n D^kT^n s(x) \qquad
\sum_n \vert c_n\vert^2 < \infty \}.
\label{wav.5}
\eeq
The resolution is determined by the width of the support of these
functions, which for $L=3$, is $5 \times 2^{-k}$.   
The functions 
\beq
s^k_n(x) := D^k T^n (x)s(x) ,
\label{wav.6}
\eeq
for fixed $k$, 
are orthonormal, have compact support on $[2^{-k}n,2^{-k}(n+5)]$,
satisfy
\beq
\int s^k_n(x) dx = 2^{-k/2}
\label{wav.7}
\eeq
and are locally finite partitions of unity
\beq
\sum_n 2^{k/2} s^k_n (x)=1 . 
\label{wav.8}
\eeq
The subspace ${\cal S}^k$ is called the resolution $2^{-k}$ subspace of
$L^2(\mathbb{R})$. 

The scale transformation $D$ has the following intertwining properties
with translations and derivatives:
\beq
T D = D T^2 \qquad \mbox{and}
\qquad {d \over dx} D = 2 D {d \over dx}.  
\label{wav.9}
\eeq
Applying $D^kT^n$ to the renormalization group equation, using
(\ref{wav.9}), gives
\beq
s^k_n (x) = \sum_{l=0}^{2L-1} h_l D^{k+1} T^{2n+l} s(x) =
\sum_{l=0}^{2L-1} h_l s^{k+1}_{2n+l}(x)
\label{wav.10}
\eeq
which expresses every basis element of ${\cal S}^k$ as a finite linear
combination of basis elements of ${\cal S}^{k+1}$ or 
\beq
{\cal S}^k \subset {\cal S}^{k+1}.  
\label{wav.11}
\eeq
This means that the lower resolution subspaces are subspaces of the higher
resolution subspaces.  The orthogonal complement of ${\cal S}^{k}$
in ${\cal S}^{k+1}$ is called ${\cal W}^k$:
\beq
{\cal S}^{k+1} = {\cal S}^{k} \oplus {\cal W}^{k}.    
\label{wav.12}
\eeq
Since ${\cal W}^k \subset {\cal S}^{k+1}$, orthonormal basis
functions $w^k_n (x)$ in ${\cal W}^k$ are also linear
combinations of the $s^{k+1}_n(x)$.  These functions are defined by
\beq
w^k_n(x)=D^kT^n w(x)
\eeq
where $w(x)$ is the ``mother wavelet'' defined by 
\beq
w (x):= \sum_{l=0}^{2L-1} g_l D  T^{l} s(x) 
\label{wav.13}
\eeq
and the coefficients $g_l$ are related to the weight coefficients
$h_l$ by
\beq
g_l = (-)^l h_{2L-1-l} \qquad 0 \leq l \leq 2L-1 .
\label{wav.14}
\eeq
The orthonormal basis functions $w^k_n(x)$ for ${\cal W}^{k}$ are called
wavelets.  Since the $w^k_n(x)$ are finite linear combinations of the
$s^{k+1}_n(x)$ they have the same number of derivatives as $s (x)$.
$w^{k}_n(x)$ also has the same support as $s^{k}_n(x)$.  Finally
it follows from (\ref{wav.3}) that 
\beq
\int x^m w^k_n(x) = 0 \qquad 0 \leq m < L .
\label{wav.15}
\eeq
Equation (\ref{wav.15}) is equivalent to the condition (\ref{wav.3}).
Equation (\ref{wav.12}) means that the wavelet subspace
${\cal W}^k$ consists of
functions that increase the resolution of ${\cal S}^k$
from $2^{-k}$ to $2^{-(k+1)}$.

The inclusions (\ref{wav.11}) imply a decomposition of ${\cal S}^{k+n}$
into an orthogonal direct sum of the form
\beq
{\cal S}^{k+n} =
{\cal W}^{k+n-1}\oplus {\cal W}^{k+n-2}\oplus \cdots \oplus
{\cal W}^{k} \oplus {\cal S}^{k}
\label{wav.16}
\eeq
which indicates that the resolution of ${\cal S}^k$ can be increased
to $2^{-k-n}$ by including additional basis functions in the subspaces
$\{{\cal W}^{k+n-1}, \cdots ,{\cal W}^{k}\}$.  This can be continued
to arbitrarily fine resolutions to get all of $L^2(\mathbb{R})$:
\beq
L^2 (\mathbb{R}) = {\cal S}^{k} \oplus_{n=0}^\infty {\cal W}^{k+n} =
\oplus_{n=-\infty}^{\infty}   {\cal W}^{n}.
\label{wav.17}
\eeq  
Since all of the subspaces are orthogonal, an orthonormal basis for
$L^2 (\mathbb{R})$ consists of
\beq
\{ s^k_n(x)\}_{n=-\infty}^\infty \cup \{ w^m_n(x)\}_{n=-\infty,m=k}^\infty  
\label{wav.18}
\eeq  
for any fixed starting resolution $2^{-k}$ or 
\beq
\{ w^k_n(x)\}_{k,n=-\infty}^\infty.
\label{wav.19}
\eeq
The basis (\ref{wav.19}) includes functions of arbitrarily large
support, while the basis (\ref{wav.18}) consists of functions with
support in intervals of width $2^{-l}(2L-1)$ for $l\geq k$.

The basis (\ref{wav.18}) is used with $L=3$ Daubechies wavelets
\cite{daubechies:1988}\cite{daubechies}.  Locally finite linear
combinations of the $L=3$ scaling functions, $s^k_n(x)$, can be used
to point-wise represent polynomials of degree 2.  The wavelets,
$w^l_n(x)$, are orthogonal to these polynomials.  The $L=3$ basis
functions have one continuous derivative.


\section{wavelet representation of quantum fields}

In what follows the basis (\ref{wav.18}) is used to expand quantum
fields restricted to a light front.  It is useful to think of the
starting scale $2^{-k}$ in (\ref{wav.18}) as the resolution that is relevant to
experimental measurements.  The higher resolution degrees of freedom
are used to represent shorter distance degrees of freedom that couple
to experimental-scale degrees of freedom.

The basis (\ref{wav.18}) can be used to get a formally exact
representation of the field operators of the form
\beq
\phi (\tilde{\mathbf{x}},x^+) :=
\sum \phi_{lmn}(x^+) \xi_l (x^-) \xi_m (x^1)\xi_n (x^2)
\qquad \mbox{where} \qquad 
\phi_{lmn}(x^+)= \int {d^2x_{\perp}dx^-}
\xi_l (x^-) \xi_m (x^1)\xi_n (x^2) \phi (\tilde{\mathbf{x}},x^+)
\label{rep.1}
\eeq
and the $\xi_l$ are the basis functions 
\beq
\xi_l (x) \in \{ s^k_n(x)\}_{n=-\infty}^\infty \cup \{ w^m_n(x)\}_{n=-\infty,m=k}^\infty .
\label{rep.2}
\eeq
In what follows the short-hand notation is used
\beq
\xi_{\mathbf{n}} (\tilde{\mathbf{x}}) :=
\xi_{n_-}(x^-)\xi_{n_1}(x^1)\xi_{n_2}(x^2)
\qquad
\sum_{\mathbf{n}}= \sum_{n_-}\sum_{n_1}\sum_{n_2}.
\label{rep.3}
\eeq
With this notation (\ref{rep.1}) has the form
\beq
\phi (\tilde{\mathbf{x}},x^+) :=
\sum_{\mathbf{n}} \phi_{\mathbf{n}}(x^+) \xi_{\mathbf{n}} (\tilde{\mathbf{x}}),  
\label{rep.4}
\eeq
which gives a discrete representation of the field as a linear
combination of discrete operators with different resolutions on the
light front.

Each discrete field operator, $\phi_{\mathbf{n}}(0)$, is associated
with a degree of freedom that is localized in a given volume on the
light-front hyperplane.  In addition, there are an infinite number of
these degrees of freedom that are localized in any open set
on the light front.

While the fields are operator valued distributions, that does not
preclude the existence of operators constructed by smearing with
functions that have only one derivative.  Note that the support
condition implies that the Fourier transform of the basis functions
are entire.

\section{kinematic Poincar\'e transformations of
  fields in the wavelet representation}

Since this representation is formally exact, kinematic Poincar\'e
transformations on the algebra of fields restricted to the light-front
can be computed by acting on the basis functions.  This follows
from the kinematic covariance of the field
\beq
U(\Lambda, a) 
\phi (\tilde{\mathbf{x}},x^+=0) U^{\dagger} (\Lambda, a)   =
\phi ((\tilde{\pmb{\Lambda}}
\tilde{\mathbf{x}}+ \tilde{\mathbf{a}}),x^+=0)
\label{kpt.1}
\eeq
for $(\Lambda,a)$ in the light-front kinematic subgroup.
Using the discrete representation of the field on both sides of this
equation gives the identity 
\beq
U(\Lambda, a)
\sum_{\mathbf{n}}  \phi_{\mathbf{n}}(x^+=0)  
\xi_{\mathbf{n}} (\tilde{\mathbf{x}})U^{\dagger} (\Lambda, a)  = 
\sum_{\mathbf{n}} 
\phi_{\mathbf{n}}(x^+=0) \xi_{\mathbf{n}}
(\tilde{\pmb{\Lambda}}
\tilde{\mathbf{x}}+ \tilde{\mathbf{a}}) .  
\label{kpt.2}
\eeq
This shows that kinematic transformations can be computed exactly
by transforming
the arguments of the expansion functions.

The transformation property of the discrete field operators restricted to
a light front follows from the orthonormality of the basis functions
(\ref{kpt.2}):
\beq
U(\Lambda, a)
\phi_{\mathbf{n}}(x^+=0)  U^{\dagger} (\Lambda, a) =
\sum_{\mathbf{m}}  
\phi_{\mathbf{m}}(x^+=0) U_{\mathbf{m} \mathbf{n}}
(\tilde{\pmb{\Lambda}},\tilde{\mathbf{a}})
\label{kpt.3}
\eeq
where the matrix 
\beq
U_{\mathbf{m} \mathbf{n}}
(\tilde{\pmb{\Lambda}},\tilde{\mathbf{a}}):=
\int {d^2x_{\perp}dx^-}\xi_{\mathbf{m}}
(\tilde{\pmb{\Lambda}}
\tilde{\mathbf{x}}+ \tilde{\mathbf{a}})
\xi_{\mathbf{n}} (\tilde{\mathbf{x}})
\label{kpt.4}
\eeq
is a discrete representation of the light front kinematic subgroup.

This identity implies that in the wavelet representation
kinematic Lorentz transformations on the
fields can be computed either 
by transforming the arguments of the basis functions
or by transforming the discrete
field operators. 

\section{States in the wavelet representation}

Because the algebra of free fields restricted to the light front is
irreducible and kinematically invariant interactions
leave the Fock vacuum unchanged, the Hilbert space for the
dynamical model can be generated by applying functions of the
discrete field operators, $\phi_{\mathbf{n}}(x^+=0)$,
to the Fock vacuum. 

Smeared light-front fields can be represented in the discrete representation as
linear combinations of the discrete field operators
\beq
\phi (f,x^+=0) := 
\sum_{\mathbf{n}}
\int {d^2x_{\perp}dx^-} f(\tilde{\mathbf{x}})\xi_{\mathbf{n}} (\tilde{\mathbf{x}})
\phi_{\mathbf{n}}(x^+=0).   
\label{hs.1}
\eeq
Equation (\ref{hs.1}) can be expressed as
\beq
\phi (f,x^+=0) = 
\sum_{\mathbf{n}}
f_{\mathbf{n}}\phi_{\mathbf{n}}(x^+=0)
\label{hs.2}
\eeq
where
\beq
f_{\mathbf{n}}:=
\int {d^2x_{\perp}dx^-} f(\tilde{\mathbf{x}})\xi_{\mathbf{n}} (\tilde{\mathbf{x}}).
\label{hs.3}
\eeq   

States can be expressed as polynomials in the smeared fields applied to
the light-front Fock vacuum
\beq
\sum c_{m_1 \cdots m_n} \phi(f_{m_1},0) \cdots \phi(f_{m_n},0) \vert 0 \rangle .
\label{hs.4}
\eeq
This representation can be re-expressed as a linear combination of
products of discrete fields applied to the Fock vacuum
\beq
\sum c_{\mathbf{m}_1 \cdots \mathbf{m}_n}  \phi_{\mathbf{m}_1}( 0) \cdots
\phi_{\mathbf{m}_n}(0) \vert 0 \rangle .
\label{hs.5}
\eeq
The inner product of two vectors of this form is a linear combination
of $n$-point functions.  For the free field algebra, the $n$-point functions
are products
of two-point functions.  The two-point functions have the form 
\beq
\langle 0 \vert \phi (f,0) \phi(g,0) \vert 0 \rangle =
\int {\theta (p^+)dp^+ d^2 p_{\perp} \over 2p^+}
\tilde{f}(-\tilde{\mathbf{p}} ) \tilde{g}(\tilde{\mathbf{p}}). 
\label{hs.6}
\eeq
This integral is logarithmically divergent if the Fourier transforms
of the smearing functions do not vanish at $p^+=0$.  Since $p^+=0$
corresponds to infinite $3$-momentum, this requirement is that the
smearing functions need to vanish for infinite 3-momentum.

From (\ref{hs.2}) and (\ref{hs.6}) it follows that the inner product above is a linear
combination of two-point functions in the discrete fields,
$\phi_{\mathbf{n}}(x^+=0)$.

The basis functions $\xi_{m} (x)$ have compact support which implies that
their Fourier transforms are entire functions of the light-front momenta
$\tilde{\mathbf{p}}$.  This means that they cannot vanish in a
neighborhood of $p^+=0$, however they can have isolated zeroes at
$p^+=0$.  For the wavelet basis functions, $w^l_m(x)$, the vanishing 
(\ref{wav.15}) of the first three moments of the $L=3$ wavelets implies that  
\beq
\tilde{w}^l_m (p^+)_{p^+=0}  = {1 \over {2\pi}^{1/2}}\int w^l_m(x^-) dx^- =0
\qquad
{d \over dp^+}\tilde{w}^l_m (p^+)_{p^+=0}  = - 
{1 \over {2\pi}^{1/2}}\int x^- w^l_m(x^-) dx^- =0
\label{hs.7}
\eeq
\beq
{d^2 \over d^2p^+}\tilde{w}^l_m (p^+)_{p^+=0}  = -
{1 \over {2\pi}^{1/2}}\int (x^-)^2 w^l_m(x^-) dx^- =0 .
\label{hs.8}
\eeq
Since the Fourier transforms are entire this means that
they have the form $\tilde{w}^l_m (p^+) = (p^+)^3f_m^l (p^+)$
where $f_m^l (p^+)$ is entire.  For the scaling function 
basis functions, $s^k_m(x)$, the normalization condition (\ref{wav.8})
gives
\beq
\tilde{s}^k_m (p^+)_{p^+=0}  = {1 \over {2\pi}^{1/2}}\int s^k_m(x^-) dx^- =
{1 \over {2\pi}^{1/2}}2^{-k/2} \not=0 .
\label{hs.9}
\eeq
These results imply that
\beq
\langle 0 \vert
\phi_{\mathbf{m}}(x^+=0) \phi_{\mathbf{n}}(x^+=0) \vert 0 \rangle
\label{hs.10}
\eeq
is singular if both basis functions have scaling functions
in the $x^-$ variable, but are finite if at least one of the
basis functions has a wavelet in the $x^-$ variable.

Since the smearing functions, $f(\tilde{\mathbf{p}})$,
should all vanish at $p^+=0$, the discrete representation will
involve linear combinations of wavelets and scaling functions
whose Fourier transforms all vanish at $p^+=0$.  In computing these
quantities the linear combinations of scaling
functions should be summed before performing the integrals.  This can
alternatively be done by including a
cutoff near $p^+=0$, doing the integrals, adding the contributions and
then letting the cutoff go to zero.

\section{Dynamics}
  
The dynamical problem involves diagonalizing $P^-$ on the free field
Fock space or solving the
light-front Schr\"odinger (\ref{n.22}) or Heisenberg equations (\ref{n.23}).  The two dynamical
equations can be put in integral form
\beq
\Psi (x^+) \vert 0 \rangle  = \Psi(x^+=0)\vert 0 \rangle
-{i \over 2}\int_0^{x^+} [P^-,\Psi (x^{+\prime})] \vert 0 \rangle dx^{+\prime} 
\label{dyn.1}
\eeq
or 
\beq
O(x^+)  = O(x^+=0) + {i \over 2}\int_0^{x^+} dx^{+\prime}
[P^-,O(x^{+\prime})] 
\label{dyn.2}
\eeq
where $\Psi(x^+=0)$ and $O(x^+=0)$ are operators in the light-front
Fock algebra. 

The formal iterative solution of these equations has the structure of a linear
combination of products of discrete fields, $\phi_{\mathbf{n}}(0)$, in
the light front Fock algebra
with $x^+$-dependent coefficients.  What is needed to perform this iteration
are the initial operators $\Psi(x^+=0)$ and $O(x^+=0)$ expressed as
polynomials in the $\phi_{\mathbf{n}}(0)$, the expression for
$P^-$ as a polynomial in the $\phi_{\mathbf{n}}(0)$, and an
expression for the commutator,
$[\phi_{\mathbf{m}}(0),\phi_{\mathbf{n}}(0)]$,
of the discrete fields on the light front.

\section{The commutator}

It follows from (\ref{f.14}) that the commutator of the discrete fields is
\beq
[ \phi_{\mathbf{m}}(0),\phi_{\mathbf{n}}(0)]
=
-{i \over 4} \delta_{m_1 n_1}\delta_{m_2 n_2}
\int \xi_{m^-}(x^-) \epsilon (x^--y^-) \xi_{n^-}(y^-)
dx^- dy^- .
\label{cr.1}
\eeq
Unlike the inner product, the commutator is always finite since
both $\xi_{m^-}(x^-)$ and  $\xi_{n^-}(y^-)$ have compact support.

The commutator (\ref{cr.1}) can be computed exactly using the
renormalization group equations.  The computation involves three
steps.  The first step is to express $\xi_{m^-}(x^-)$ and
$\xi_{n^-}(y^-)$ as linear combinations of scaling functions on a
sufficiently fine common scale.  The second step is to change
variables so the commutator is expressed as a linear combination of
commutators involving integer translates of the fixed-point 
solution $s(x^-)$ of the renormalization group equation.
The last step is to use the renormalization group equation to
construct a finite linear system relating the commutators involving
integer translates of the $s(x^-)$.

Applying $D^kT^n $ to the
renormalization group equation and the expression for $w(x)$ gives
\beq D^k T^n s(x) = \sum_{L=0}^l h_l D^{k+1}T^{2n+l}s(x).
\label{cr.2}
\eeq
and
\beq
D^k T^n w(x) =
\sum_{L=0}^l g_l D^{k+1}T^{2n+l}s(x).
\label{cr.3}
\eeq
These equations express  $s^k_n(x)$ and $w^k_n(x)$ as linear combinations of
the $s^{k+1}_n(x)$  :
\beq
s^k_n(x) = \sum_{l=0}^{2L-1} h_l s^{k+1}_{2n+l}(x)=
\sum_{m=2n}^{2n+2L-1} h_{m-2n} s^{k+1}_{m}(x)=
\sum_{m=2n}^{2n+2L-1} H_{n;m} s^{k+1}_{m}(x)  \qquad
\mbox{where}
\qquad
H_{n;m}:=h_{m-2n}
\label{cr.4}
\eeq
and
\beq
w^k_n (x) =\sum_{l=0}^{2L-1} g_l s^{k+1}_{2n+l}(x)=
\sum_{m=2n}^{2n+2L-1} g_{m-2n} s^{k+1}_{m} (x)= \sum_{m=2n}^{2n+2L-1} G_{n;m} s^{k+1}_{m} (x) \qquad
\mbox{where}
\qquad  G_{n;m}:=g_{m-2n}.
\label{cr.5}
\eeq
While the matrices $H_{n;m}$ and $G_{n;m}$ are formally infinite,
for each fixed $n$ these are 0 unless $2n \leq m \leq 2L-1+2n$.

Using powers of the matrices
\beq
H^m_{nl} := \sum H_{nk_1}H_{k_1k_2} \cdots H_{k_ml} 
\label{cr.6}
\eeq
and $G_{nl}$ the basis function can be represented as
finite linear combinations of finer resolution scaling functions
\beq
s^k_n = \sum_l H^m_{nl} s^{k+m}_l
\label{cr.7}
\eeq
\beq
w^k_n = \sum_{lt} H^{m-1}_{nt}G_{tl} s^{k+m}_l   
\label{cr.8}
\eeq
where the sums in (\ref{cr.7}) and (\ref{cr.8}) are finite.
Using these identities all of the integrals can be reduced to
finite linear combinations of integrals involving a pair of scaling functions,
$s^k_n(x)= 2^{k/2} s(2^kx -n)$, on a common fine scale, $2^{-k}$.

What remains is linear combinations of products of integrals of the form
\[
\int s^k_m(x) \epsilon (x^--y^-) s^k_n (y^-)
dx^- dy^- =
\]
\beq
\int 2^{k/2} s(2^kx^- - m) \epsilon (x^--y^-) 2^{k/2} s(2^k y^- -n)
dx^- dy^- .
\label{cr.9}
\eeq
Changing variables 
\beq
y^{-\prime} = 2^k y^- - n, \quad  x^{-\prime} = 2^k x^- - n 
\label{cr.10}
\eeq
noting
\beq
\epsilon (x^--y^-) = \epsilon (2^{k}x^--2^{k}y^-)
\label{cr.11}
\eeq
this becomes
\beq
\int 2^{-k} s(x^{\prime-}-m) \epsilon (x^{\prime -}-y^{\prime -}) s(y^{\prime-}-n)
dx^{\prime -} dy^{\prime -} =
\label{cr.12}
\eeq
\beq
\int 2^{-k} s(x^{\prime-}+n-m) \epsilon (x^{\prime -}-y^{\prime -}) s(y^{\prime-})
dx^{\prime -} dy^{\prime -} =
2^{-k} I[n-m]
\label{cr.13}
\eeq
where
\beq
I[n]:=\int s(x^{-} + n) \epsilon (x^--y^{-}) s(y^{-})
dx^{-} dy^{-}.
\label{cr.14}
\eeq
$I[n]$ can be expressed as a difference of two integrals
\beq
I[n]=\int s(x^{-}+n ) [\int_{-\infty}^{x^- } s(y^{-})
- \int_{x^- }^\infty s(y^-) ]
dx^{-} dy^{-} 
\label{cr.15}
\eeq
while the normalization condition (\ref{wav.4}) gives
\beq
\int s(x^{-}+n  ) [\int_{-\infty}^{x^-  } s(y^{-})
+ \int_{x^- }^\infty s(y^-) ]
dx^{-} dy^{-} =1 .
\label{cr.16}
\eeq
Adding (\ref{cr.15}) and (\ref{cr.16})  gives:
\beq
I[n]= 2 \int
s(x^{-}+n ) \int_{-\infty}^{x^-} s(y^{-})dx^{-} dy^{-} -1 .
\label{cr.17}
\eeq
If the support of $s(x^{-}+n )$ is to the right of the
support of $s(y^-)$, the integral is 1 while if the
support of $s(x^{-}+n )$ is to the left of the
support of $s(y^-)$ the integral is $-1$.
Thus for the $L=3$ basis functions
\beq 
I[n]  =
\left \{
\begin{array}{c l}
1 & n \leq =-5\\    
I[n]  & -4 \leq n \leq 4\\  
-1 &n \geq 5 \\   
\end{array}
\right . .
\label{cr.18}
\eeq
The $I[n]$ for $n \in [-4,4]$ are related by the renormalization group
equations
\beq
I[n]=\int s(x^{-}+n ) \epsilon (x^--y^- )s(y^-) 
dx^{-} dy^{-} =
\label{cr.19}
\eeq
\beq
2 \sum h_l h_k
\int s(2x^{-}+2n-l ) \epsilon (x^--y^- )s(2y^--k) 
  dx^{-} dy^{-} =
\label{cr.20}
\eeq
\beq
{1 \over 2} \sum h_l h_k
\int s(2x^{-}+2n-l ) \epsilon (2x^--2y^- )s(2y^--k) 
  2dx^{-} 2dy^{-} =
\label{cr.21}
\eeq
\beq
{1 \over 2} \sum h_l h_k
\int s(x^{-}+2n-l ) \epsilon (x^- -y^- )s(y^--k) 
dx^{-} dy^{-} =
\label{cr.22}
\eeq
\beq
{1 \over 2} \sum h_l h_k
\int s(x^{-}+2n-l+k ) \epsilon (x^- -y^- )s(y^-) 
dx^{-} dy^{-} =  
\label{cr.23}
\eeq
\beq
{1 \over 2} \sum h_l h_k I[2n+k-l] =
{1 \over 2} \sum h_{m+l-2n} h_l I[m] =
{1 \over 4} \sum a_{m-2n}I[m]
\label{cr.26}
\eeq
where
\beq
a_n:=2 \sum_{l=0}^5 h_l h_{l+n}  \qquad -5 \leq n \leq 5 .
\label{cr.27}
\eeq

The numbers $a_n$ will appear again.
The $a_n$ are rational numbers
\cite{beylkin1}\cite{beylkin2}\cite{beylkin-92}\cite{beylkin3}.  For L=3 the non-zero
$a_n$ are 
\beq
a_0=2 \qquad
a_1=a_{-1}={75\over 64} \qquad
a_3=a_{-3}=-{25\over 128} \qquad
a_5=a_{-5}={3\over 128}. 
\label{cr.28}
\eeq
The $9 \times 9$ matrix $A_{mn} := a_{n-2m}$ ($-4\leq m,n \leq 4$)
has the following rational eigenvalues
$\lambda =2,1,{1 \over 2}, {1 \over 4}, \pm {1\over 8}, {1 \over 16},
{9 \over 32}, -{9\over 64}$, so it is invertible.

The non-trivial $I[n]$ are solutions of the linear system
\beq
\sum_{n=-4}^4 A_{mn} I[n] = d_m
\label{cr.29}
\eeq
where
\beq
d_m =a_{5-2m}-a_{-5-2m}.
\label{cr.30}
\eeq
The solution of (\ref{cr.29}) is 
\beq
I[n]=
\left (
\begin{array}{r l}
-3.34201389e+00,&n=-4\\
 8.33333333e+00,&n=-3\\
-1.79796007e+01,&n=-2\\
 1.94444444e+01,&n=-1\\
 0.00000000e-00,&n=0\\
-1.94444444e+01,&n=1\\
 1.79796007e+01,&n=2\\
-8.33333333e+00,&n=3\\
 3.34201389e+00,&n=4\\
\end{array}
\right )  .
\label{cr.31}
\eeq
While (\ref{cr.31}) is a numerical solution,
the exact solution is rational since both $A_{mn}$ and $d_n$ are rational.

This solution, along with (\ref{cr.18}), can be used to construct the commutator
of any of the discrete field operators using (\ref{cr.2}-\ref{cr.12}).

The general structure of the commutators is
\beq
    [ \phi_{\mathbf{m}}(0),\phi_{\mathbf{n}}(0)] =
    C_{\mathbf{m},\mathbf{n}}=
(\mbox{scale factors}) \times (\mbox{powers of H,G}) \times I[n] 
\eeq

Note that while this commutator looks very non-local, if the scaling
functions in (\ref{cr.1}) are replaced by wavelets with supports
that are sufficiently separated,  the integrals vanish because
the moments of wavelets vanish.  This will also be true of linear
combinations of scaling functions that represent functions that vanish
at $p^+=0$.


\section{Poincar\'e generators}

The other quantity needed to formulate the dynamics is an expression
for $P^-$ or one of the other dynamical Poincar\'e generators
expressed in terms of operators in the irreducible algebra.  Since the
generators are conserved Noether charges, they are independent of
$x^+$, so the generators can be expressed in terms of fields on the
light front.  The discrete representations of the generators can be
constructed by replacing the fields on the light front by the discrete
representation (\ref{rep.1},\ref{rep.4}) of the fields.  The integrals over the light
front become integrals over products of basis functions and their
derivatives.  This section discusses the computation of these
integrals using renormalization group methods.

A scalar $\phi^4(x)$ theory is used for the purpose of illustration.
In this case the problem is to express all of the generators as linear
combinations of products of discrete fields.

The construction of the Poincar\'e generators from Noether's theorem
was given in section IV.  Using the discrete representation of fields
the light-front Poincar\'e generators (\ref{fd.25}-\ref{fd.31})
have the following forms
\beq
P^+ = \sum_{\mathbf{m}\mathbf{n}} :\phi_{\mathbf{m}} (0) \phi_{\mathbf{n}}(0):
P^+_{\mathbf{m},{\mathbf{n}}}
\label{pg.1}
\eeq
where
\beq
P^+_{\mathbf{m},{\mathbf{n}}} :=
2   
\int dx^- d^2x_{\perp} \partial_-  \xi_{\mathbf{m}}(\tilde{\mathbf{x}}) \partial_- \xi_{\mathbf{n}}(\tilde{\mathbf{x}})),  
\label{pg.2}
\eeq
\beq
P^i = \sum_{\mathbf{m}\mathbf{n}} :\phi_{\mathbf{m}} (0) \phi_{\mathbf{n}}(0):
P^i_{\mathbf{m},{\mathbf{n}}}
\label{pg.3}
\eeq
where
\beq
P^i_{\mathbf{m},{\mathbf{n}}}:=
-   
\int dx^- d^2x_{\perp}  \partial_-  \xi_{\mathbf{n}}(\tilde{\mathbf{x}}) \partial_i \xi_{\mathbf{m}}(\tilde{\mathbf{x}}),
\label{pg.4}
\eeq
\beq
P^{-} =
\sum_{\mathbf{m}\mathbf{n}} :\phi_{\mathbf{m}} (0) \phi_{\mathbf{n}}(0):
P^-_{\mathbf{m},\mathbf{n}} +
\sum_{\mathbf{n}_1\mathbf{n}_2 \mathbf{n}_4\mathbf{n}_4}
:\phi_{\mathbf{n}_1} (0) \phi_{\mathbf{n}_2}(0)\phi_{\mathbf{n}_3} (0)
\phi_{\mathbf{n}_4}(0):
P^-_{\mathbf{n}_1,\mathbf{n}_2, \mathbf{n}_3,\mathbf{n}_4}
\label{pg.5}
\eeq
where
\beq
P^-_{\mathbf{m},\mathbf{n}}:=
\int dx^- d^2x_{\perp}  \left ( {1 \over 2}
\pmb{\nabla}_{\perp}\xi_{\mathbf{m}}(\tilde{\mathbf{x}})\cdot\pmb{\nabla}_{\perp}\xi_{\mathbf{n}}(\tilde{\mathbf{x}})
+ {1 \over 2}m^2 
\xi_{\mathbf{m}}(\tilde{\mathbf{x}})\xi_{\mathbf{n}}(\tilde{\mathbf{x}})\right )
\label{pg.6}
\eeq
and
\beq
P^-_{\mathbf{n}_1,\mathbf{n}_2, \mathbf{n}_3,\mathbf{n}_4} :=
\lambda
\int dx^- d^2x_{\perp}
\xi_{\mathbf{n}_1}(\tilde{\mathbf{x}})\xi_{\mathbf{n}_2}(\tilde{\mathbf{x}})
\xi_{\mathbf{n}_3}(\tilde{\mathbf{x}})\xi_{\mathbf{n}_4}(\tilde{\mathbf{x}}),
\label{pg.7}
\eeq
\beq
K^3 =
\sum_{\mathbf{m}\mathbf{n}} :\phi_{\mathbf{m}} (0) \phi_{\mathbf{n}}(0):
K^3_{\mathbf{m},\mathbf{n}} +
\sum_{\mathbf{n}_1\mathbf{n}_2 \mathbf{n}_4\mathbf{n}_4}
:\phi_{\mathbf{n}_1} (0) \phi_{\mathbf{n}_2}(0)\phi_{\mathbf{n}_3} (0)
\phi_{\mathbf{n}_4}(0):
K^3_{\mathbf{n}_1,\mathbf{n}_2, \mathbf{n}_3,\mathbf{n}_4}
\label{pg.8}
\eeq
where
\beq
K^3_{\mathbf{m},\mathbf{n}} :=
\int dx^- d^2 \left (  2x_{\perp} x^- \partial_-  \xi_{\mathbf{m}}(\tilde{\mathbf{x}}) \partial_- \xi_{\mathbf{n}}(\tilde{\mathbf{x}})
- {1 \over 2} x^+ \pmb{\nabla}_{\perp}\xi_{\mathbf{m}}(\tilde{\mathbf{x}})\cdot\pmb{\nabla}_{\perp}\xi_{\mathbf{n}}(\tilde{\mathbf{x}})
-{1 \over 2} m^2  x^+ \xi_{\mathbf{m}}(\tilde{\mathbf{x}})\xi_{\mathbf{n}}(\tilde{\mathbf{x}})\right )
\label{pg.9}
\eeq
and
\beq
K^3_{\mathbf{n}_1,\mathbf{n}_2, \mathbf{n}_3,\mathbf{n}_4}:=
-{\lambda} 
\int dx^- d^2x_{\perp}
x^+ \xi_{\mathbf{n}_1}(\tilde{\mathbf{x}})\xi_{\mathbf{n}_2}(\tilde{\mathbf{x}})\xi_{\mathbf{n}_3}(\tilde{\mathbf{x}})\xi_{\mathbf{n}_4}(\tilde{\mathbf{x}}).
\label{pg.10}
\eeq
Setting $x^+=0$ this becomes
\beq
K^3_{\mathbf{m},\mathbf{n}} \to
2 \int dx^- d^2x_{\perp} x^- \partial_-  \xi_{\mathbf{m}}(\tilde{\mathbf{x}}) \partial_- \xi_{\mathbf{n}}(\tilde{\mathbf{x}});
\qquad
K^3_{\mathbf{n}_1,\mathbf{n}_2, \mathbf{n}_3,\mathbf{n}_4} \to 0 .
\label{pg.11}
\eeq
For the remaining generators
\beq
E^1 =
\sum_{\mathbf{m}\mathbf{n}} :\phi_{\mathbf{m}} (0) \phi_{\mathbf{n}}(0):
E^1_{\mathbf{m},\mathbf{n}}
\label{pg.12}
\eeq
where 
\beq
E^1_{\mathbf{m},\mathbf{n}} :=
\int dx^- d^2x_{\perp} \left (
2 x^1 \partial_- \xi_{\mathbf{m}}(\tilde{\mathbf{x}}) \partial_- \xi_{\mathbf{n}}(\tilde{\mathbf{x}}) + \partial_-\xi_{\mathbf{m}}(\tilde{\mathbf{x}}) \partial_1\xi_{\mathbf{n}}(\tilde{\mathbf{x}})x^+\right )  \to
2\int x^1 \partial_- \xi_{\mathbf{m}}(\tilde{\mathbf{x}}) \partial_- \xi_{\mathbf{n}}(\tilde{\mathbf{x}}),
\label{pg.13}
\eeq
\beq
E^2 = \sum_{\mathbf{m}\mathbf{n}} :\phi_{\mathbf{m}} (0) \phi_{\mathbf{n}}(0):
E^2_{\mathbf{m},\mathbf{n}}
\label{pg.14}
\eeq
where
\beq
E^2_{\mathbf{m},\mathbf{n}} :=
\int dx^- d^2x_{\perp} (
2 x^2 \partial_- \xi_{\mathbf{m}}(\tilde{\mathbf{x}}) \partial_- \xi_{\mathbf{n}}(\tilde{\mathbf{x}}) + \partial_-\xi_{\mathbf{m}}(\tilde{\mathbf{x}}) \partial_2\xi_{\mathbf{n}}(\tilde{\mathbf{x}})x^+  )\to
2\int dx^- d^2x_{\perp} x^2 \partial_- \xi_{\mathbf{m}}(\tilde{\mathbf{x}}) \partial_- \xi_{\mathbf{n}}(\tilde{\mathbf{x}}), 
\label{pg.15}
\eeq
\beq
F^1=
\sum_{\mathbf{m}\mathbf{n}} :\phi_{\mathbf{m}} (0) \phi_{\mathbf{n}}(0):
F^1_{\mathbf{m},\mathbf{n}} +
\sum_{\mathbf{n}_1\mathbf{n}_2 \mathbf{n}_3\mathbf{n}_4}
:\phi_{\mathbf{n}_1} (0) \phi_{\mathbf{n}_2}(0)\phi_{\mathbf{n}_3} (0)
\phi_{\mathbf{n}_4}(0):
F^1_{\mathbf{n}_1,\mathbf{n}_2, \mathbf{n}_3,\mathbf{n}_4}
\label{pg.16}
\eeq
where
\beq
F^1_{\mathbf{m},\mathbf{n}} :=
\int dx^- d^2x_{\perp} \left ({1 \over 2} x^1 \pmb{\nabla}_{\perp}\xi_{\mathbf{k}}(x)\cdot\pmb{\nabla}_{\perp}\xi_{\mathbf{l}}(x) + 
{1 \over 2}x^1 m^2 \xi_{\mathbf{k}}(\tilde{\mathbf{x}}) \xi_{\mathbf{l}}(\tilde{\mathbf{x}})+  x^- \partial_- \xi_{\mathbf{k}}(\tilde{\mathbf{x}})\partial_1 \xi_{\mathbf{l}}(\tilde{\mathbf{x}})\right )
\label{pg.17}
\eeq
and
\beq
F^1_{\mathbf{n}_1,\mathbf{n}_2, \mathbf{n}_3,\mathbf{n}_4} :=
\lambda \int dx^- d^2x_{\perp}
x^1 \xi_{\mathbf{n}_1}(\tilde{\mathbf{x}})
\xi_{\mathbf{n}_2}(\tilde{\mathbf{x}})\xi_{\mathbf{n}_3}(\tilde{\mathbf{x}})\xi_{\mathbf{n}_4}(\tilde{\mathbf{x}}),
\label{pg.18}
\eeq
\beq
F^2=
\sum_{\mathbf{m}\mathbf{n}} :\phi_{\mathbf{m}} (0) \phi_{\mathbf{n}}(0):
F^2_{\mathbf{m},\mathbf{n}} +
\sum_{\mathbf{n}_1\mathbf{n}_2 \mathbf{n}_4\mathbf{n}_4}
:\phi_{\mathbf{n}_1} (0) \phi_{\mathbf{n}_2}(0)\phi_{\mathbf{n}_3} (0)
\phi_{\mathbf{n}_4}(0):
F^2_{\mathbf{n}_1,\mathbf{n}_2, \mathbf{n}_3,\mathbf{n}_4}
\label{pg.19}
\eeq
where
\beq
F^2_{\mathbf{m},\mathbf{n}} :=
\int dx^- d^2x_{\perp} \left ({1 \over 2} x^2 \pmb{\nabla}_{\perp}\xi_{\mathbf{k}}(x)\cdot\pmb{\nabla}_{\perp}\xi_{\mathbf{l}}(x) + 
{1 \over 2}x^2 m^2 \xi_{\mathbf{k}}(\tilde{\mathbf{x}}) \xi_{\mathbf{l}}(\tilde{\mathbf{x}})+  x^- \partial_- \xi_{\mathbf{k}}(\tilde{\mathbf{x}})\partial_2 \xi_{\mathbf{l}}(\tilde{\mathbf{x}})\right ) 
\label{pg.20}
\eeq
and
\beq
F^2_{\mathbf{n}_1,\mathbf{n}_2, \mathbf{n}_3,\mathbf{n}_4} :=
\lambda \int dx^- d^2x_{\perp}
x^2 \xi_{\mathbf{n}_1}(\tilde{\mathbf{x}})
\xi_{\mathbf{n}_2}(\tilde{\mathbf{x}})\xi_{\mathbf{n}_3}(\tilde{\mathbf{x}})\xi_{\mathbf{n}_4}(\tilde{\mathbf{x}}).
\label{pg.21}
\eeq
All of these operators have the structure of linear combinations of
normal products of discrete fields evaluated at $x^+=0$ times constant
coefficients, 
$
P^+_{\mathbf{n}_1,\mathbf{n}_2}, 
P^i_{\mathbf{n}_1,\mathbf{n}_2}, 
P^-_{\mathbf{n}_1,\mathbf{n}_2},
P^-_{\mathbf{n}_1,\mathbf{n}_2, \mathbf{n}_3,\mathbf{n}_4}, 
K^3_{\mathbf{n}_1,\mathbf{n}_2}, 
J^3_{\mathbf{n}_1,\mathbf{n}_2}, 
E^i_{\mathbf{n}_1,\mathbf{n}_2}, 
F^i_{\mathbf{n}_1,\mathbf{n}_2}, 
F^i_{\mathbf{n}_1,\mathbf{n}_2, \mathbf{n}_3,\mathbf{n}_4} 
$,
which are integrals
involving products of basis
functions and their derivatives.  The three-dimensional integrals that
need to be evaluated to compute these coefficients are products of
three one-dimensional integrals that have one of the following eight
forms:
\beq
\int dx \xi_{{m}}({x}) \xi_{{n}}(x)
\qquad
\int dx \partial_x  \xi_{{m}}(x)  \xi_{{n}}(x)
\qquad
\int dx \partial_x  \xi_{{m}}(x) \partial_x \xi_{{n}}(x)
\label{pg.22}
\eeq
\beq
\qquad
\int dx x \xi_{{m}}(x) \xi_{{n}}(x)
\qquad
\int dx x \partial_x  \xi_{{m}}(x)  \xi_{{n}}(x)
\qquad
\int dx x \partial_x  \xi_{{m}}(x) \partial_x \xi_{{n}}(x) .
\label{pg.23}
\eeq
\beq
\int dx  \xi_{{n}_1}(x) \xi_{{n}_2}(x) \xi_{{n}_3}(x)\xi_{{n}_4}(x)
\qquad
\int dx x \xi_{{n}_1}(x) \xi_{{n}_2}(x) \xi_{{n}_3}(x)\xi_{{n}_4}(x).
\label{pg.24}
\eeq
In what follows it is shown how all of these integrals
can be computed using the renormalization group equation (\ref{wav.1}).

The integrals (\ref{pg.22}-\ref{pg.24}) are products of basis
functions which may be scaling functions with scale $2^{-k}$ or
wavelets of scale $2^{-k-l}$ for $l \geq 0$.
The same methods that were used in the computation of the commutator
function, (\ref{cr.2}-\ref{cr.8}), can be used to express the integrals (\ref{pg.22}-\ref{pg.24}) as linear combinations
of integrals involving scaling functions on a common scale fine scale,
$2^{-l}$.

After expressing the integrals in terms of
scaling functions, $s^l_n(x)$, and their derivatives,
the one-dimensional integrals
(\ref{pg.22}-\ref{pg.24}) can be expressed in terms of integrals
involving products of the $s_n(x)$.  A variable change, $x \to x' =
2^{-l}x$ can be used to express all of the integrals in terms of
translates of the original fixed point $s(x)$.  The scale factors
for each type of integral are shown below:
\beq
\int dx s^l_{{m}}(x) s^l_{{n}}(x) = \delta_{mn}
\label{pg.25}
\eeq
\beq
\int dx \partial_x s^l_{{m}}(x)  s^l_{{n}}(x) =
2^l \int dx s^{\prime} (x)  s_{{n-m}}(x)
\label{pg.26}
\eeq
\beq
\int dx \partial_x  s^l_{m}(x) \partial_x s^l_{n}(x) =
2^{2l} \int dx s' (x)  s'_{n-m}(x)
\label{pg.27}
\eeq
\beq
\int dx  s^l_{n_1}(x) s^l_{n_2}(x) s^l_{n_3}(x)s^l_{n_4}(x)=
2^l \int dx s(x) s_{n_2-n_1}(x) s_{n_3-n_1}(x)s_{n_4-n_1}(x)
\label{pg.28}
\eeq
\beq
\int dx x  s^l_{{m}}(x)  s^l_{{n}}(x) =
2^{-l} (\int dx x s (x)  s_{{n-m}}(x) + m\delta_{m,n})
\label{pg.29}
\eeq
\beq
\int dx x\partial_x  s^l_{{m}}(x) s^l_{{n}}(x) =
\int dx (x+m) s^{\prime} (x)  s_{{n-m}}(x)
\label{pg.30}
\eeq
\beq
\int dx x \partial_x  s^l_{{m}}(x) \partial_x s^l_{{n}}(x) =
2^{l} \int dx (x+m) s^{\prime} (x)  s'_{{n-m}}(x)
\label{pg.31}
\eeq
\beq
\int dx  x s^l_{{n}_1}(x) s^l_{{n}_2}(x) s^l_{{n}_3}(x)s^l_{{n}_4}(x)=
\int dx (x+n_1) s(x) s_{{n}_2-n_1}(x) s_{{n}_3-n_1}(x)s_{{n}_4-n_1}(x).
\label{pg.32}
\eeq
These identities express all of the integrals involving scale $2^{-l}$
scaling functions in terms of related integrals involving the
$s_n(x)$.  The compact support of the functions $s_n(x)$ means the
these integrals are identically zero unless the indices and the
absolute values of their differences are less than $2L-2$ which is $4$
for $L=3$.

The integrals of the right side of (\ref{pg.26}-\ref{pg.32}),
are the following integrals:
\beq
\delta_{mn} = \int dx s_{{m}}(x) s_{{n}}(x)  \qquad m=n
\label{pg.33}
\eeq

\beq
D_1[m] := \int dx {ds_{{}}\over dx} (x) s_{{m}}(x) \qquad -4 \leq m \leq 4
\label{pg.34}
\eeq

\beq
D_2[m] := \int dx {d s_{{}}\over dx}(x) {ds_{{m}}\over dx}(x) \qquad -4 \leq m \leq 4
\label{pg.35}
\eeq

\beq
\Gamma_4[m][n][k] :=
\int dx s(x) s_{m}(x) s_{{n}}(x)s_{{k}}(x)
\qquad -4 \leq m,n,k,m-n,m-k,k-n \leq 4
\label{pg.36}
\eeq

\beq
X[m] := \int dx x s_{{}}(x) s_{{m}}(x) \qquad -4 \leq m \leq 4
\label{pg.37}
\eeq

\beq
X_1[m] := \int dx x {ds_{{}}\over dx} (x) s_{{m}}(x) \qquad -4 \leq m \leq 4
\label{pg.38}
\eeq

\beq
X_{2}[m] := \int dx x {d s_{{}}\over dx}(x) {ds_{{m}}\over dx}(x) \qquad -4 \leq m \leq 4
\label{pg.39}
\eeq

\beq
\Gamma_{4x}[m][n][k] :=
\int dx x s(x) s_{m}(x) s_{{n}}(x)s_{{k}}(x)
\qquad -4 \leq m,n,k,m-n,m-k,k-n \leq 4 .
\label{pg.40}
\eeq
The renormalization group equation in the form
\beq
s(x-n) = \sum_{l=0}^{5} h_l \sqrt{2} s( 2x-2n -l)
\label{pg.41}
\eeq
and a variable change $x \to x' = 2x$  leads to the following linear equations
relating the non-zero values of these integrals
\beq
D_1[n] = \sum_{m=-4}^4 a_{m-2n} D_1[m] = \sum_{m=-4}^4 A_{nm} D_1[m]
\label{pg.42}
\eeq
\beq
D_2[n] = 2\sum_{m=-4}^4 a_{m-2n} D_2[m] = 2 \sum_{m=-4}^4 A_{nm} D_2[m]
\label{pg.43}
\eeq
where $a_m$
\beq
a_{m} := 2 \sum_{k=0}^{5} h_{k+m} h_k   \qquad -5 \leq m \leq 5
\label{pg.43.1}
\eeq
is the same quantity (\ref{cr.27}-\ref{cr.28}) that appeared in the
computation of the commutator function.
A similar quantity appears in the homogeneous
equations relating the non-zero $\Gamma_4[m][n][k]$'s: 
\[
\Gamma_4[m][n][k] := \sum_{l,l_m l_n,l_k=0}^{5}
2 h_{l}h_{l_m}h_{l_n}h_{l_k}  
\Gamma_4[2m+l_m-l][2n+l_n-l][2k+l_k-l] =
\]
\beq
\sum_{m'm'k'} A_4 (m,n,k;m',n',k') \Gamma_4[m'][n'][k']
\label{pg.44}
\eeq
where
\beq
A_4 (m,n,k;m',n',k') := \sum_l 2 h_l h_{m'-2m+l}h_{n'-2n+l}h_{k'-2k+l}. 
\label{pg.44.1}
\eeq
The relations involving $X[n],X1[n],X2[n]$ and $\Gamma_{4x}[m][n][k]$
have inhomogeneous parts
\beq
X[n]= {1 \over 4} \sum_{m=-4}^4 A_{nm} X[m] +
{1 \over 2} \sum_{l} l h_l h_{l-2n}  
\label{pg.45}
\eeq
\beq
X_1[n]= {1 \over 2} \sum_{m=-4}^4 A_{nm} X_1[m] +
\sum_{l} l h_l h_{l-2n+m}D_1[m]   
\label{pg.46}
\eeq
\beq
X_2[n]=  \sum_{m=-4}^4  A_{nm} X_2[m] +
2 \sum_{l} l h_l h_{l-2n+m}D_2[m]   
\label{pg.47}
\eeq
\beq
\Gamma_{4x}[m][n][k] :=
{1 \over 2} \sum_{m'n'k'} A_4(m,n,k;m',n',k')
\Gamma_{4x}[m'][n'][k'] -
\label{pg.48}
\eeq
\beq
\sum_{m'n'k'}(\sum_l
h_{l}h_{m'-2m+l}h_{n'-2n+l}h_{k'-2k+l}l)
\Gamma_{4}[m'][n'][k'].
\label{pg.49}
\eeq
Since the $9 \times 9$ matrix $A_{mn} := a_{n-2m}$ ($-4\leq m,n \leq 4$)
has eigenvalues
$\lambda =2,1,{1 \over 2}, {1 \over 4}, \pm {1\over 8}, {1 \over 16},
{9 \over 32}, -{9\over 64}$,  it follows that $D_1[n]$ and $D_2[n]$
are eigenvectors of $A_{mn}$ with eigenvalues $1$ and ${1\over 2}$
respectively.  The normalization is determined by the
equations discussed below.  Equation (\ref{pg.44}) similarly
implies that $\Gamma_4[m][n][k]$ is an eigenvector with eigenvalue 1 of the
matrix $A_4$ defined by the right-hand side of (\ref{pg.44}).
The normalization  of $\Gamma_4[m][n][k]$ is also discussed below.

The matrix $(I-{1 \over 4}A)$ in (\ref{pg.45}) is invertible so
(\ref{pg.45}) is a well-posed linear system for $X[n]$, while the
matrices $(I-{1 \over 2}A)$ and $(I-A)$ in (\ref{pg.46}) and
(\ref{pg.47}) are singular.  To solve them the Moore-Penrose
generalized inverse \cite{benisrael} is applied to the inhomogeneous
terms to get specific solutions.  These solutions are
substituted back in the equations to ensure that the inhomogeneous
terms are in the range of $(I-{1 \over  2}A)$ and $(I-A)$ respectively,
although this must be the case since the solutions can also be expressed
as integrals.   The general solutions of
({\ref{pg.46}) and (\ref{pg.47}) can include arbitrary amounts of the
solution of the homogeneous equations which are eigenstates of
$A_{mn}$ with eigenvalues 2 and 1 respectively.  The contribution
from the homogeneous equation is determined by
the normalization conditions below.

The normalization conditions are derived from the
property that polynomials with degree less than $L$ can be point-wise
represented as locally finite-linear combination of the $s_n(x)$.
These expansions have the form
\beq
1 = \sum s_n (x)
\label{pg.51}
\eeq
\beq
x = \sum (\langle x \rangle + n) s_n(x) = \langle x \rangle + \sum n s_n(x) 
\label{pg.52}
\eeq
\beq
x^2 = \sum (\langle x \rangle + n)^2 s_n(x) =
\langle x \rangle^2 + 2\langle x \rangle \sum n s_n(x) + \sum n^2 s_n(x)  .
\label{pg.53}
\eeq
where
\beq
\langle x^n \rangle := \int s(x) x^n dx
\eeq
are moments of $s(x)$.
Differentiating (\ref{pg.52}) and (\ref{pg.53})
gives
\beq
1 = \sum n s'_n(x) 
\label{pg.54}
\eeq
\beq
x =
\langle x \rangle + {1 \over 2} \sum n^2 s'_n(x)  .
\label{pg.55}
\eeq
Multiplying (\ref{pg.54}) by $s(x)$ and integrating the result
gives
\beq
\sum_{n=-4}^4  n D_1[n] = -1 .
\label{pg.58}
\eeq
Multiplying (\ref{pg.55}) by $s'(x)$ and integrating gives
\beq
\sum_{n=-4}^4  n^2 D_2[n] = -2. 
\label{pg.59}
\eeq
These conditions determine the normalization of the eigenvectors $D_1[n]$ and
$D_2[n]$. Note that the moments do not appear
in these normalization conditions,  although all moments of $s(x)$
can be
computed recursively using renormalization group equation and the normalization
condition (\ref{wav.4}).
Using (\ref{pg.54}) in (\ref{pg.38}) and integrating by parts gives:
\beq
\sum_{n=-4}^4 X_1[n] = -1 .
\label{pg.61}
\eeq
Using (\ref{pg.55}) in (\ref{pg.39}) and integrating by parts gives:
\beq
\sum_{n=-4}^4 n X_2[n] = -1 . 
\label{pg.62}
\eeq
These conditions determine the contribution of the solution of the
homogeneous equations in the general solution.

The normalization conditions for $\Gamma_4[m][n][k]$ are obtained
using the partition of unity property (\ref{pg.51})
\beq
\sum_{m=-4}^4 \Gamma_4[m][n][k] = \Gamma_3[n][k];
\qquad
\sum_{n=-4}^4 \Gamma_3[n][k] = \delta_{k0}
\label{pg.63}
\eeq
\beq
\sum_{m=-4}^4 \Gamma_{4x}[m][n][k] = \Gamma_{3x}[n][k];
\qquad
\sum_{n=-4}^4 \Gamma_3[n][k] = X[k]
\label{pg.64}
\eeq
where 
\beq
\Gamma_{3}[m][n] :=
\int dx x s(x) s_{m}(x) s_{{n}}(x)
\qquad -2L+2 \leq m,n,m-n \leq 2L-2
\label{pg.65}
\eeq
\beq
\Gamma_{3x}[m][n] :=
\int dx x s(x) s_{m}(x) s_{{n}}(x)
\qquad -2L+2 \leq m,n,m-n \leq 2L-2
\label{pg.66}
\eeq
and $\Gamma_3[m][n]$ is a solution of the eigenvalue problem
\beq
\Gamma_3[m][n] =
\sum_{m'n'} a_3(m,n;m'n') \Gamma_3[m'][n']
\label{pg.67}
\eeq
with normalization (\ref{pg.63}) and
\beq
a_3(m,n;m'n') = \sum_l h_l h_{m'-2m+l}h_{n'-2n+l} .
\eeq
$\Gamma_{3x}[m][n]$ satisfies 
\beq
\Gamma_{3x}[m][n] :=  \sum_{m'n'} a_3(m,n;m'n') \Gamma_{3x}[m'][n']- 
\label{pg.68}
\eeq
\beq
\sum_{m'n'}(
\sum_{l}^{2L-1}
lh_l h_{m'-2m+l}h_{n'-2n+l})
\Gamma_{3}[m'][n']
\label{pg.69}
\eeq
with the normalization constraint 
\beq
\sum_n \Gamma_{3x}[m][n] = X[m].
\label{pg.69.1}
\eeq
These finite linear systems can be solved for all of the integrals
(\ref{pg.22}-\ref{pg.24}).  The results for
$D1[n],D2[n],X[n],X1[n],X2[n]$ for $L=3$, which are needed to compute
the constant coefficients for the free field generators are given
below.  The vector $\Gamma_4[m][n][k]$ of coefficients for
the dynamical generators has too many components to display.
They can be computed by finding the
eigenvector with eigenvalue 1 of the $9^3 \times 9^3$ matrix
$a_4[m][n][m'][n']$ with normalization given by (\ref{pg.63}).  The
normalization condition requires solving for the eigenvector with
eigenvalues 1 of the $9^2 \times 9^2$ matrix $a_3[m][n][m'][n']$ using
the normalization condition (\ref{pg.63}).  Finally
$\Gamma_{4x}[m][n][k]$ be computed by applying the Moore Penrose
generalized inverse of $(I-a_4)$ to the inhomogeneous term in (\ref{pg.48})
and adding an amount of the solution of the eigenvalue problem
$(2I-a_4)X=0$ consistent with the normalization condition (\ref{pg.64})

All the these quantities can alternatively computed by a direct quadrature,
however the fractal nature of the basis functions makes the
renormalization group method discussed above preferable.  The values of
$D1[n], D2[n], X[n], X1[n]$ and $X2[n]$ are given below:
\beq
\left (
\begin{array}{l}
D1[-4]={1 \over 2920}\\    
D1[-3]=-{16 \over 1095}\\      
D1[-2]=-{53 \over 365}\\      
D1[-1]={272 \over 365}\\      
D1[0]=0.0\\
D1[1]=-{272 \over 365}\\
D1[2]={53 \over 365}\\
D1[3]=-{16 \over 1095}\\
D1[4]=-{1 \over 2920}\\
\end{array}
\right )
\qquad
\left (
\begin{array}{l}
D2[-4]=-{3 \over 560}\\    
D2[-3]=-{4 \over 35}\\      
D2[-2]={92 \over 105}\\      
D2[-1]=-{356 \over 105}\\      
D2[0]={295 \over 56}\\
D2[1]=-{356 \over 105}\\
D2[2]={92 \over 105}\\
D2[3]=-{4 \over 35}\\
D2[4]=-{3 \over 560}\\
\end{array}
\right )
\eeq
\beq
\left (
\begin{array}{l}
X0[-4]=-3.96222254e-06\\    
X0[-3]=-6.76219313e-04\\      
X0[-2]=1.92128831e-02\\      
X0[-1]=-1.21043257e-01\\      
X0[0]=  1.02242228e+00\\
X0[1]= -1.21043257e-01\\
X0[2]=  1.92128831e-02\\
X0[3]= -6.76219313e-04\\
X0[4]= -3.96222254e-06\\
\end{array}
\right )
\qquad
\left (
\begin{array}{l}
X1[-4]= 1.75026831e-06\\
X1[-3]=-6.81293512e-04\\
X1[-2]=-3.98947081e-02\\
X1[-1]=3.39841948e-01\\
X1[0]=-5.00000000e-01\\
X1[1]=-1.08504743e+00\\
X1[2]=3.30305667e-01\\
X1[3]=-4.31543229e-02\\
X1[4]= -1.37161328e-03\\
\end{array}
\right )
\qquad 
\left (
\begin{array}{l}  
X2[-4]=-5.08087952e-04\\
X2[-3]=-8.68468406e-03\\
X2[-2]=5.47476157e-01\\
X2[-1]=-3.01673853e+00\\
X2[0]=6.95730703e+00\\
X2[1]=-6.40481025e+00\\
X2[2]=2.29938859e+00\\
X2[3]=-3.51494681e-01\\
X2[4]= -2.19355544e-02\\
\end{array}
\right )
\eeq
\section{truncations}

The value of the wavelet representation is that, while it is formally exact,
it also admits natural volume and resolution truncations in the
light-front hyperplane.  Truncations define effective theories that
are expected to be good approximations to the theory for reactions
associated with a volume and energy scale corresponding to the volume
and resolution of the truncations.  The simplest truncation discards
degrees of freedom smaller than some limiting fine resolution,
$2^{-l}$ as well as degrees of freedom with support outside of some
volume on the light front.

In this regard it has similar properties to a lattice truncation.
Unlike a lattice truncation, because the theory is formally exact it
is straightforward to systematically include corrections associated
with finer resolution or larger volumes.  Some other appealing
features are that the truncated fields have a continuous space-time
dependence and can be differentiated, so there is no need to use
finite difference approximations.  Finally it is possible to take
advantage of some of the advantages of the light-front quantization.

One problem that is common to lattice truncations of field theory is
that truncations break symmetries.  In the light-front case
truncations break the kinematic covariance.  One consequence is that
transforming the truncated field covariantly using (\ref{kpt.1}) is
not the same as transforming the truncated field using the matrix
(\ref{kpt.3}) and truncating the result.  The difference between these
two calculations is due to the discarded degrees of freedom, which
should be small for a suitable truncation.  This suggests that
kinematic Lorentz transformations can be approximated by using
(\ref{kpt.1}) with the truncated fields.  The vacuum of the formally
exact theory is the trivial Fock vacuum if the interaction commutes
with the kinematic subgroup.  When the kinematic invariance is broken
the lowest mass eigenstate of the truncated $P^-$ is not necessarily
the Fock vacuum, however the Fock vacuum states should become the
lowest mass state in the infinite-volume, zero-resolution limit.  This
suggests that using trivial Fock vacuum might still be a good
approximation.

The basis discussed in this work is not the only possible basis choice
and may not be the best option for treating the transverse degrees of
freedom for fields in $3+1$ dimensions.  In this work the transverse
degrees of freedom are expanded in products of multi-scale basis
functions of Cartesian coordinates, $x$ and $y$.  Truncations of this
basis break the rotational symmetry about the $z$ axis. An alternative
is to expand the transverse degrees of freedom in a basis consisting
of products of functions of the polar coordinates $r$ and $\theta$
where $x=r\cos(\theta)$ and $y=r \sin (\theta)$.  The basis functions
in the 
$\theta$ variable can be taken as the periodic functions, ${1 \over
  \sqrt{2\pi}}e^{i n\theta}$.  This choice maintains the rotational
symmetry, but does not give a multi-resolution treatment of the angle
degree of freedom.  A second option is to use the multi-resolution
basis in the angle variable on $[0,2\pi]$ with periodic boundary
conditions.  In this case the truncations will result in a discrete
rotational symmetry that depends on the resolution.  In both cases the
radial degree of freedom can be expanded in a multi-resolution basis.
The only difference is that the radial functions have support on
$[0,\infty]$ rather than $[-\infty,\infty]$.  This requires replacing
the basis functions that have support at $r=0$ by linear combinations
of these functions that satisfy the boundary conditions at the origin.
The linear combinations in a subspace of a given resolution can be
constructed so they are orthonormal on $[0,\infty]$, resulting in an
orthonormal basis on that subspace, however the modified basis
functions near the origin in subspaces of different resolution are no
longer orthogonal.  This results in additional coupling of degrees of
freedom on different scales near $r=0$. This is because the exact
boundary conditions at $r=0$ involve functions of all resolutions.

\section{Summary and outlook}

This work introduced a multi-resolution representation of quantum
field theory on a light front.  This is a formally exact
representation of the field theory in terms of an infinite number of
discrete degrees of freedom that are localized on the light front.
Each degree of freedom is associated with a compact subset of the
light front.  These subsets cover the light front and there are an
infinite number of them in every open subset on the light front.  This
representation has the property that there are a finite number of
these degrees of freedom associated with any finite volume and any
given maximal resolution on the light front.

Each degree of freedom or mode is represented by a field on the light
front integrated over a basis function of compact support on the light
front.  The discrete fields associated with a free-field theory are an
irreducible set of operators on the free field Fock space.  For
interacting theories with self-adjoint kinematically invariant
interactions the spectral condition on $P^+$ implies that the
interaction cannot change the Fock vacuum.  This means dynamical
operators like the Poincar\'e generators can be expressed as functions
of this irreducible algebra of fields acting on the free field Fock
space.

The Poincar\'e generators involve ill-defined products of fields at the same
point, so the formal interactions are not well-defined
self-adjoint operators on the Fock space.  In the multi-resolution
representation the ultraviolet singularities that arise from local
operator products necessarily appear as non-convergence of infinite
sums of well-defined operator products.  There are also infrared
divergences that appear in products of scaling function modes even
after the smearing.  In the light-front case the ultraviolet and
infrared singularities are constrained by rotational covariance, so
any strategy to non-perturbatively renormalize the theory must treat
these problems together. 

Computations necessarily involve both volume and resolution cutoffs,
which result in a well-defined truncated theory with a finite number
of degrees of freedom.  As long as the interaction in the truncated
theory vanishes at $p^+=0$, the interaction will leave the Fock
vacuum unchanged.  The variable $p^+ = \hat{\mathbf{z}}\cdot \mathbf{p} + \sqrt{m^2 + \mathbf{p}^2}$ approaches zero in the limit that
$-\hat{\mathbf{z}}\cdot \mathbf{p} \to +\infty$, so it is an infinite
momentum limit, which involves high-resolution degrees of freedom.
Requiring that the interaction vanish at $p^+=0$ is
a resolution cutoff.  This can be realized by discarding
products of scaling function modes in the interaction.  These modes
do not contribute to the operator product when it is integrated over
functions with vanishing Fourier transforms at $p^+=0$.

Dynamical calculations evolve the fields to points off of the light
front.  This evolution can be performed by iterating the light-front
Heisenberg field equations or by solving the light-front Schrodinger
equation.  Both cases involve discrete mathematics.  Iterating the
Heisenberg field equations results in a representation of the field as
an expansion in normal products of discrete fields on the light front
with $x^+$-dependent coefficients.  Because fields on the light front
are irreducible, the different discrete field operators cannot all
commute, however the commutator can be calculated explicitly and
analytically.  Vacuum expectation values of product of fields can be
computed by evaluating the solution of the Heisenberg equations in the
Fock vacuum.

There are a number of problems involving free fields that can be used
to try to understand the convergence of computational strategies in
truncated theories.  Free field theories have the advantage that they
can be solved exactly, so errors can be calculated by comparing exact
computations to computations based on truncated theories.  Among the
problems of interest is how is Poincar\'e invariance recovered as the
resolution is increased in a truncated theory.  Because the basis is
local, this can be tested in a finite volume.  Methods for performing
this test in the corresponding wavelet representation of canonical
field theory were discussed in \cite{fatih2}.  These methods utilize
the locally finite partition of unity property of the scaling
functions in the expression for the generators in terms of the
integrals over the energy-momentum and angular-momentum tensor
densities.  In the light-front case free fields provide a laboratory
to investigate the accuracy of kinematic Lorentz transformations in
truncated theories.  Another important problem is how efficiently can
the multi-resolution representation of the light front Hamiltonian be
block diagonalized by resolution.  This was studied for the case of
the corresponding wavelet representation of canonical field theory in
\cite{tracie}.  One conclusion of that work is that both volume and
resolution need to be increased simultaneously in order to converge to
a sensible energy spectrum of the Hamiltonian (i.e so it approaches a
continuous spectrum that is unbounded above).  In addition, it was
found that convergence to a block diagonal form slowed as energy
separation of the modes decreased.  Another calculation that should be
done is to compare the Wightman functions or commutator functions of
the truncated theories to the exact quantities.  The light front
representation has the advantage that these can be computed without
without solving for an approximate vacuum.  Another interesting
question is what is the contribution of the product of the infrared
singular parts of the truncated fields to normal ordered products of
free fields.  Does the normal ordering remove these contributions?

The next class of problems of interest are 1+1 dimensional solvable
field theories.  These are interesting because the dynamical equations
in the multi-resolution representation generates more complicated
operators in the algebra of fields on the light front.  Reference
\cite{best:1994} used Daubechies' wavelets methods in a canonical
representation of the field theory to treat the X-Y model and
spontaneous symmetry breaking in the Landau Ginzburg model.

The real interest is to apply multi-resolution methods to realistic
theories in 3+1 dimensions.  These are computationally far more
complex than problems involving free fields or problem in low
dimensions.  There are several kinds of problems of interest.  These
include bound state problems, scattering problems, studies of
correlation functions and extensions to gauge theories.  While the
discrete nature of the multi-resolution representation has some
computational advantages, they will not be of significant help for
these complex problems, especially since the number of modes scale
with dimension and number of particles.  One of the advantages of
multi-resolution methods is that basis functions are self similar.  The
result is that the coupling strength of the various modes
differs by different powers of 2. A systematic investigation could
help to identify the most dominant modes in a given application.  This
could be used to get a rough first approximation and can be improved
perturbatively.  One interesting property of the multi-resolution
representation of the theory is that it is both discrete and formally
exact. In a formally exact theory Haag-Ruelle scattering theory can be
used to express scattering observables as strong limits.  Of interest
is to use the exact representation to develop an approximation
algorithm for computing scattering observables in this discrete
representation.  This is not trivial, since the time limits will not
converge if they are computed after truncation.

Bound state calculations could be computed by diagonalizing the
mass operator on a subspace, similar to how this is done using
basis light front quantization \cite{james}.  Variational methods
could prove useful in this regard.

For gauge theories, the exact representation of the field theory in
terms of a countable number of discrete fields with different
resolutions suggest that a similar construction could be performed in
using of gauge invariant degrees of freedom.  To understand how this
might work imagine a set of gauge invariant Wilson placquets with a
given lattice spacing on a light front.  The expectation is that on
the light front these form an irreducible algebra of operators of a
given resolution.  Decreasing the lattice spacing by a factor of 2,
results in a new algebra that is an irreducible set of operators for
the increased resolution.  The coarse-scale algebra should be a sub
algebra of the fine-scale algebra.  In the same way that scaling
functions on a file scale can be expressed as wavelet and scaling
functions on a coarse scale, is can be anticipated that there is
something like a wavelet transform that generates the fine scale
algebra in terms of generators for the coarse scale algebra and
independent gauge invariant operators that generate the degrees of
freedom in the fine scale algebra that are not in the coarse scale
algebra.  As in the wavelet case, this could be repeated on every
scale, leading to a countable set of independent operators that can
generate placquets on all scales.  This should result in an
irreducible set of gauge invariant operators on the light front,
with a formally trivial vacuum.  While this construction is far from
trivial, having a formally exact representation of gauge theories in
terms of local gauge invariant variables is a desirable goal.

Another class of applications where the wavelet representation may be
useful is in quantum computing.  The fundamental property is that the
local nature of the interactions involving different discrete modes
means that transfer matrices for small time steps can be expressed as
simple quantum circuits.  Some comments on using wavelet discretized
fields in quantum computing appear in references \cite{Brennen} and
\cite{Evenbly}.  The advantage in the light front case is the trivial
nature of the vacuum.

The author would like to acknowledge generous support from the
U.S. Department of Energy, Office of Science, Nuclear
Theory Program, Grant number DE-SC16457.

\end{document}